\documentclass[twocolumn]{aastex62}

\usepackage{amssymb} 
\usepackage{amsmath} 
\usepackage{graphicx}
\usepackage{latexsym}
\usepackage{dcolumn}
\usepackage{bm}

\usepackage{color}

\newcommand{\be}{\begin{equation}}
\newcommand{\ee}{\end{equation}}
\newcommand{\ba}{\begin{eqnarray}}
\newcommand{\ea}{\end{eqnarray}}

\newcommand{\bfi}{\begin{figure}
\epsfxsize=8cm
\epsffile}
\newcommand{\bfig}{\begin{figure*}
\epsfxsize=18cm
\epsffile}
\newcommand{\efi}{\end{figure}}
\newcommand{\efig}{\end{figure*}}
\newcommand{\bi}{\begin{itemize}}
\newcommand{\ei}{\end{itemize}}

\newcommand{\mpch}{h^{-1} {\rm Mpc}}

\newcommand{\hmpc}{h {\rm Mpc}^{-1}}

\graphicspath{{./}{figures/}}

\received{January 1, 2018}
\revised{January 7, 2018}
\accepted{\today}
\submitjournal{ApJ}

\shorttitle{NR of Vel}
\shortauthors{Yu et al.}

\begin{document}

\title{Nonlinear Reconstruction of the Velocity Field}

\author[0000-0002-9359-7170]{Yu Yu}
\affiliation{Department of Astronomy, Shanghai Jiao Tong University,
800 Dongchuan Road, Shanghai 200240, China}

\author[0000-0002-8202-8642]{Hong-Ming Zhu}
\affiliation{Berkeley Center for Cosmological Physics and Department of Physics,
University of California, Berkeley, CA 94720, USA}
\affiliation{Lawrence Berkeley National Laboratory, 1 Cyclotron Road, Berkeley, CA 94720, USA}

\begin{abstract}
We propose a new velocity reconstruction method based on the displacement estimation of recently developed methods.
The velocity is first reconstructed by transfer functions in Lagrangian space and then mapped into Eulerian space.
High-resolution simulations are used to test the performance.
We find that the new reconstruction method outperforms the standard velocity reconstruction in the sense of better cross-correlation coefficient, less velocity misalignment, and smaller amplitude difference.
We conclude that this new method has the potential to improve the large-scale structure sciences involving a velocity reconstruction, such as kinetic Sunyaev--Zel'dovich measurement and supernova cosmology. 
\end{abstract}



\section{Introduction}
\label{sec:Introduction}

Reconstructing the velocity field from density is a non-trivial procedure due to the high nonlinearity in the evolved density field and the non-local relationship with the velocity field.
Also, usually we only have discrete and biased tracers such as galaxies in the observation, which suffer from density bias, shot noise, and stochasticity.
Early attempts, e.g. \cite{Nusser94}, \cite{Fisher95}, etc., aimed to estimate the peculiar velocities from the early pioneering galaxy surveys.
These results improved our knowledge of the Local Universe.

Discovering the accelerating expansion of our universe using supernovas observation is a triumph of modern cosmology.
This discovery utilizes the luminosity distance and redshift relation to constrain the cosmological models.
The Doppler effect by the peculiar velocity is one of the systematics in the supernova cosmology.
Neglecting correlated peculiar velocities can cause an error in the best-fit value of the dark energy equation of state
and also an overestimate of the precision of the measurement (\cite{Cooray06,Hui06,Davis11}).
The low-redshift cutoff is usually applied in order to avoid this systematics.
For future Supernova surveys attempting statistical error bars of less than about $2\%$, it is important to correct the peculiar velocity.
The reconstructed velocity field is also helpful for reducing the Hubble constant measurement uncertainty from standard sirens (\cite{Mukherjee19}).

The kinetic Sunyaev--Zel'dovich (kSZ) effect offers a unique opportunity to characterize the cosmic peculiar velocity field in the distant universe, and to search for the ‘missing baryons.’
The kSZ measurement benefits from the velocity estimates for avoiding the cancellation of equally likely positive and negative kSZ signals (\cite{DeDeo05,Ho09,Shao11a,Li14a,Smith18}).
The significance level depends on the velocity estimation/reconstruction performance.
In recent years, works have detected and measured the kSZ signal. 
\cite{Hand12} first reported the detection of the kSZ signal by applying the pairwise kSZ estimator to ACT cosmic microwave background (CMB) data using a galaxy catalog from the Sloan Digital Sky Survey (SDSS) III DR9.
This measurement was achieved with higher precision using the Baryon Oscillation Spectroscopic Survey (BOSS) DR11 catalog (\cite{De-Bernardis17}).
With the Planck CMB map, \cite{Planck-kSZ16} reported a kSZ detection using the Central Galaxy Catalog extracted from SDSS DR7 and \cite{Li18} presented the measurement using BOSS data.
Most of the velocity reconstruction methods used in the above literature are motivated by the linearized continuity equation.
One can solve for the velocity field from the observed density field with some preprocessing such as debiasing, smoothing, redshift-space distortion (RSD) correction, and Gaussianization.
The reconstruction is performed in Eulerian space
and only the irrotational part is reconstructed by design.
For future CMB-S4 surveys and Stage-IV galaxy surveys, remote dipole and quadrupole reconstruction from the kSZ effect will benefit from a precisely reconstructed velocity field (\cite{Cayuso18,Deutsch18,Pan19,McCarthy19}).

Some physical quantities in Lagrangian space suffer less nonlinear effects and they provide us alternative angles to study the behavior of our universe.
The evolved density field is highly nonlinear in Eulerian space.
Part of the nonlinearity comes from the large-scale bulk motion, which dominates the displacement field in Lagrangian picture.
The removal of the bulk motion, i.e. the density reconstruction, recovers the linear information and sharpens the baryon acoustic oscillation peaks (e.g.,\cite{Eisenstein07}).
The velocity field is also dominated by the large-scale bulk flow.
Thus, the displacement and the velocity field are expected to correlate well in Lagrangian space.
By investigating their relation between the them in Lagrangian space, one can develop new velocity reconstruction methods.
Exploring how to reconstruct the velocity in Lagrangian space is worthwhile given that the displacement can be well estimated from the nonlinear density field.

Reconstruction of the early state of our universe has a long history.
Early achievements were accomplished by finding the least-action solution, fast action method, solving the Monge-Amp\'ere equation (e.g., \cite{Peebles89,Croft97,Branchini02,Brenier03}), and etc.
However, the performance was limited by the calculating ability.
\cite{Lavaux08}, \cite{Lavaux08a} converted the reconstructed displacement to the velocity field with the linear relation, $\boldsymbol{v}=\beta\boldsymbol{\Psi}$, with the linear growth factor $\beta\approx\Omega_m^{5/9}$.

Recently, new algorithms were proposed to reconstruct the initial condition from the highly nonlinear density map, which improves the signal-to-noise in the measurement of the baryon acoustic oscillation sound horizon scale (\cite{Zhu17,Schmittfull17,Shi18,Hada18,Sarpa19}).
The performances on the biased tracers such as the simulated halos/HOD galaxy samples are tested in \cite{Yu17b}, \cite{Birkin19}, and \cite{Hada19}.
Despite the different theoretical motivations and operational procedures in these backward modeling studies, the key to the improvement is the same, a better estimate of the nonlinear displacement.
Note that the initial condition/displacement could also be obtained from the forward modeling methods such as Hamiltonian Markov Chain Monte Carlo method (\cite{Wang14b}), optimization with a Bayesian approach (\cite{Seljak17,Modi18,Schmidt19,Jasche19}), and etc.

Note that 
the reconstruction of the displacement also inspires many potential applications.
The reconstructed displacement is an effective displacement that ensures the correct clustering but ignores some complicated processes like shell-crossing.
Understanding the reconstructed displacement could help us develop a fast mock generation method.
Given the nonlinear density field with an RSD effect, the reconstructed one also contains RSD information and this may improve the RSD modeling since the RSD is more linear after the reconstruction (\cite{Zhu18a}).
The reconstructed displacement is also useful to measure the relative velocity of the neutrino to DM, which contains important information on the neutrino mass (\cite{Zhu19}).
The reconstructed displacement also helps with moving the observable in Eulerian space back to its Lagrangian position where it is more physically positioned (such as the angular momentum of the galaxy, \cite{YuHR19}).

This paper is an investigation of velocity reconstruction using recently proposed displacement reconstruction methods.
In Section \ref{sec:motivation}, the theoretical bases of the standard velocity reconstruction method and the new proposed velocity reconstruction are introduced, and the algorithm is presented.
In Section \ref{sec:results}, we present the performance of the new velocity reconstruction.
Section \ref{sec:conclusion} summarizes the results and provides a discussion.
Additional dimensions of observations, such as the shot noise and stochasticity for biased tracers, survey mask and RSD effect, are out of the scope of this paper and will be addressed in future investigations.

\section{Motivation}
\label{sec:motivation}

\subsection{Standard velocity reconstruction}
\label{sec:standard}

The standard velocity reconstruction method adopts the linearized continuity equation,
\begin{equation}
\label{eqn:continuity}
\frac{\partial \delta(\boldsymbol{x})}{\partial t}+\nabla \cdot \boldsymbol{v}(\boldsymbol{x})=0\ ,
\end{equation}
 to convert the density maps into velocity maps.
 The reconstructed velocity field is obtained by the relation in Fourier space:
\begin{equation}
\label{eqn:stdrec}
\boldsymbol{v}_{r}(\boldsymbol{k})=afH\frac{i \boldsymbol{k}}{k^{2}} \frac{\delta_{S}(\boldsymbol{k})}{b}\ ,
\end{equation}
in which Gaussian smoothing is usually adopted to reduce the impact from the highly non-Gaussian region, linearizes the field,
and galaxy bias is corrected.
The prefactor $afH$ comes from the linear theory. $H$ is the Hubble parameter, $f=d\ln D/d\ln a$, and $D$ is the linear growth rate.
Throughout the paper, we denote the reconstructed velocity with a subscript $r$ and the true velocity field is labeled with a subscript $t$.

For the purposes of velocity reconstruction, widely adopted Gaussian smoothing might not be optimal.
There exist other schemes trying to achieve better performance.
For example, one can linearize the density field by a logarithmic transform, and/or 
obtain the velocity field using second-order Lagrangian perturbation theory (see \cite{Planck-kSZ16} for a detailed velocity reconstruction comparison for the purposes of kSZ measurement).

Here we extend the standard velocity reconstruction formalism to use the transfer function to ensure that the process is under an optimal:
\begin{equation}
\label{eqn:stdrec2}
\boldsymbol{v}_{r}(\boldsymbol{k})=\frac{i \boldsymbol{k}}{k^{2}}T(k)\delta(\boldsymbol{k})\ .
\end{equation}
The transfer function is defined to minimize the error in the reconstruction,
and is calibrated from the simulation.
Hereafter, we refer to this process as the standard reconstruction.

\subsection{Reconstruction in Lagrangian space}
\label{sec:lpt}

In the Lagrangian scenario, the motion of the fluid element is labeled by its original position $\boldsymbol{q}$ and specified by the displacement $\boldsymbol{\Psi}(\boldsymbol{q},t)=\boldsymbol{x}(t)-\boldsymbol{q}$ at time $t$.
The Lagrangian perturbation theory attempts to model the nonlinear displacement in a perturbative way,
\begin{equation}
\boldsymbol{\Psi}=\mathbf{\Psi}^{(1)}+\mathbf{\Psi}^{(2)}+\cdots\ ,
\end{equation}
in which each term collects the contribution from the same order,
\begin{equation}
\begin{aligned} \mathbf{\Psi}^{(n)}(\boldsymbol{k})=& \frac{i D^{n}}{n !} \int \frac{d^{3} k_{1}}{(2 \pi)^{3}} \cdots \frac{d^{3} k_{n}}{(2 \pi)^{3}}(2 \pi)^{3} \delta^{D}\left(\sum_{j=1}^{n} \boldsymbol{k}_{j}-\boldsymbol{k}\right) \\ & \times \boldsymbol{L}^{(n)}\left(\boldsymbol{k}_{1}, \ldots, \boldsymbol{k}_{n}\right) \delta_{L}\left(\boldsymbol{k}_{1}\right) \cdots \delta_{L}\left(\boldsymbol{k}_{n}\right)\ , \end{aligned}
\end{equation}
with $\boldsymbol{k}=\boldsymbol{k}_{1}+\cdots+\boldsymbol{k}_{n}$.
The first-order and second-order kernels are given by
\begin{equation}
\boldsymbol{L}^{(1)}\left(\boldsymbol{k}_{1}\right)=\frac{\boldsymbol{k}}{k_{1}}\ ,
\end{equation}
\begin{equation}
\boldsymbol{L}^{(2)}\left(\boldsymbol{k}_{1}, \boldsymbol{k}_{2}\right)=\frac{3}{7} \frac{\boldsymbol{k}}{k^{2}}\left[1-\left(\frac{\boldsymbol{k}_{1} \cdot \boldsymbol{k}_{2}}{k_{1} k_{2}}\right)^{2}\right]\ .
\end{equation}
The displacement divergence $\delta_{\psi}=-\nabla\cdot\boldsymbol{\Psi}$ is given by
\begin{equation}
\delta_{\psi}(\boldsymbol{k})=\delta^{(1)}(\boldsymbol{k})+\delta^{(2)}(\boldsymbol{k})+\cdots\ ,
\end{equation}
where the first order is just the linear density field, $\delta^{(1)}(\boldsymbol{k})=\delta_{L}(\boldsymbol{k})$,
and 
\begin{equation}
\begin{aligned} \delta^{(2)}(\boldsymbol{k})=& \frac{1}{7} \int \frac{d^{3} k_{1}}{(2 \pi)^{3}} \frac{d^{3} k_{2}}{(2 \pi)^{3}}(2 \pi)^{3} \delta^{D}\left(\boldsymbol{k}_{1}+\boldsymbol{k}_{2}-\boldsymbol{k}\right) \\ & \times\left\{1-\frac{3}{2}\left[\left(\frac{\boldsymbol{k}_{1} \cdot \boldsymbol{k}_{2}}{k_{1} k_{2}}\right)^{2}-\frac{1}{3}\right]\right\} \delta_L(\boldsymbol{k}_{1}) \delta_L\boldsymbol{k}_{2})\ , \end{aligned}
\end{equation}
or equivalently in configuration space,
\begin{equation}
\label{eqn:lptorder}
\delta^{(2)}(\boldsymbol{q})=\frac{1}{7} \delta_{L}^{2}(\boldsymbol{q})-\frac{1}{7} K^{2}(\boldsymbol{q})\ .
\end{equation}
The tidal term $K^2(\boldsymbol{q})$ is give by the contraction of the tidal tensor,
\begin{equation}
K^{2}(\boldsymbol{q})=\frac{3}{2} K_{i j}(\boldsymbol{q}) K_{i j}(\boldsymbol{q})\ ,
\end{equation}
where 
\begin{equation}
K_{i j}(\boldsymbol{k})=\left(\frac{k_{i} k_{j}}{k^{2}}-\frac{1}{3} \delta_{i j}\right) \delta_{L}(\boldsymbol{k})\ .
\end{equation}
Note that both $\boldsymbol{\Psi}(\boldsymbol{q})$ and $\delta_{\psi}(\boldsymbol{q})$ are in Lagrangian configuration space.


The velocity is the time derivative of the displacement.
Thus, the Lagrangian velocity is the summation of the contributions from all orders,
\begin{equation}
\boldsymbol{v}(\boldsymbol{q})=a \dot{\boldsymbol{\Psi}}(\boldsymbol{q})=a f H \boldsymbol{\Psi}^{(1)}+2 afH \boldsymbol{\Psi}^{(2)}+\cdots\ ,
\end{equation}
in which the prefactors in each term come from the linear theory for $\boldsymbol{v}$ and $\boldsymbol{\Psi}$.
In the general case, the velocity divergence is related to densities by a series of transfer functions at each order,
\begin{equation}
\label{eqn:nlrec}
\theta(\boldsymbol{k})=T_{1}(k) \delta_{L}(\boldsymbol{k})+T_{2}(k) \delta^{(2)}(\boldsymbol{k})+\cdots\ .
\end{equation}

The basic idea is that
once the linear density field is estimated by the reconstruction algorithm from the observed density field, one can use Eq. \ref{eqn:nlrec} to obtain the Lagrangian velocity field and further map it into Eulerian space.
Note that the reconstruction of the linear density is not perfect.
It contains noise and non-Gaussianity induced by the reconstruction algorithm.
When high-order terms in Eq. \ref{eqn:nlrec} are adopted,
we need to further process (described below) on the reconstructed linear density field to calculate the high-order terms.

\section{Implementation}
\label{sec:implementation}

\subsection{Simulation setup}
\label{sec:simulation}

To test and compare the velocity reconstruction methods,
we use a high-resolution simulation involving $2048^3$ dark matter particles in a box with a side length of $600\ \mpch$. It is run by the particle-particle-particle-mesh $N$-body simulation code \texttt{CUBEP$^3$M} (see \cite{Harnois-Deraps13a}).
The cosmic velocity field has a large correlation scale, typically $\sim 150 \mpch$.
This simulation box size is insufficient for robust large-scale velocity statistics measurement.
However, the following results are mainly based on the cross-correlation analysis.
Due to the cancellation of the sample variance, it is sufficient to obtain reliable results.
The reconstruction and analysis are performed on $512^3$ grids.
We assign particle velocity onto uniform grids by the nearest particle (NP) method.
Sampling artifacts in the E-mode power spectrum measurement associated with the NP assignment can be neglected in this configuration since the number density of $\sim 1 (\mpch)^{-3}$ is sufficiently high (see \cite{Zhang15,Zheng15}).

\subsection{Reconstruction algorithm}
\label{sec:algorithm}

We use three recently developed nonlinear reconstruction algorithms in this work.
They are described in \cite{Zhu17}, \cite{Shi16} and \cite{Schmittfull17}.
We denote them as A1, A2, and A3, respectively.
They all provide the reconstructed density field, which has significantly improved correlation with the linear initial condition.
Although the performance in recovering the cross-correlation coefficient is similar, these three independently developed procedures produce different behaviors in the reconstructed density field.
This leads to slightly different performances for the velocity reconstruction.
In the main result below we only show the result from A2 and present the difference in the Appendix.

Once we obtain the reconstructed density field, based on Eq. \ref{eqn:nlrec}, we propose the direct velocity reconstruction in Lagrangian space by only using the first-order term ($\mathcal{O}(1)$ reconstruction),
\begin{equation}
\theta_r(\boldsymbol{k})=T_{1}(k) \delta_{r}(\boldsymbol{k})\ .
\end{equation}
Here $\delta_r$ is the reconstructed linear density field and
the transfer function is defined as
\begin{equation}
T_1(k)=\frac{\langle\delta_r\theta_t\rangle}{\langle\delta_r\delta_r\rangle}\ .
\end{equation}
It is calibrated by the reconstructed density $\delta_r(\boldsymbol{q})$ and the true Lagrangian velocity divergence $\theta_t(\boldsymbol{q})$ in simulation.\\

The reconstructed displacement contains nonlinear information.
The high-order terms may contain useful information for reconstructing the nonlinear velocity field.
Similar to \cite{Schmittfull17}, we propose the $\mathcal{O}(2)$ reconstruction by  taking the second term in Eq. \ref{eqn:nlrec} into further consideration.

First, for the estimated linear density field $\delta_r$, we use a Wiener filter to remove the spurious power induced by the reconstruction algorithm:
\begin{eqnarray}
\begin{aligned}
&W_\mathrm{WF}(k)=\frac{\langle\delta_L\delta_r\rangle}{\langle\delta_r\delta_r\rangle}\ ,\\
&\delta_r^{(1)}(\boldsymbol{k})=\delta_r(\boldsymbol{k})W_\mathrm{WF}(k)\ .
\end{aligned}
\end{eqnarray}
Note that the reconstructed density fields from different algorithms have different noise powers, and thus different Wiener filters.
For a given algorithm, the corresponding Wiener filter should be calibrated from the mocks.
The uncertainties in the mock construction may induce uncertainties in this calibration.
However, in Wiener filtering the transit from the signal-dominating area to the noise-dominating area is rapid.
We do not expect that the uncertainties in the Wiener filtering to significantly  influence the performance.

The second-order term $\delta_r^{(2)}$ is calculated as Eq. \ref{eqn:lptorder} by replacing $\delta_L$ with $\delta_r^{(1)}$.
However, due to the residual non-Gaussianity in $\delta_r^{(1)}$, $\langle\delta_r^{(1)}\delta_r^{(2)}\rangle$ is non-zero.
Thus, we cannot directly perform second-order reconstruction based on $\delta_r^{(2)}$, otherwise the reconstructed $\theta_r^{(1)}$ and $\theta_r^{(2)}$ are not independent.
We use orthogonization technique (\cite{Schmittfull19}) to remove the correlated part in $\delta_r^{(2)}$ and construct $\hat\delta_r^{(2)}$ which has no correlation with $\delta_r^{(1)}$,
\begin{eqnarray}
\begin{aligned}
&W_\perp(k)=\frac{\langle\delta_r^{(2)}\delta_r^{(1)}\rangle}{\langle\delta_r^{(1)}\delta_r^{(1)}\rangle}\ ,\\
&\hat\delta_r^{(2)}(\boldsymbol{k})=\delta_r^{(2)}(\boldsymbol{k})-\delta_r^{(1)}(\boldsymbol{k})W_\perp(k)\ .
\end{aligned}
\end{eqnarray}
Then we can perform the first-order and second-order velocity reconstruction in sequence.
For the first-order term,
\begin{eqnarray}
\begin{aligned}
&T_1(k)=\frac{\langle\delta_r^{(1)}\theta_t\rangle}{\langle\delta_r^{(1)}\delta_r^{(1)}\rangle}\\
&\theta_r^{(1)}(\boldsymbol{k})=T_1(k)\delta_r^{(1)}(\boldsymbol{k})\ .
\end{aligned}
\end{eqnarray}
The residual is 
\begin{equation}
\theta_{m}(\boldsymbol{k})=\theta_t(\boldsymbol{k})-\theta_r^{(1)}(\boldsymbol{k})\ ,
\end{equation}
which ensures $\langle\theta_{m}\theta_r^{(1)}\rangle=0$.
For the second-order term,
\begin{eqnarray}
\begin{aligned}
&T_2(k)=\frac{\langle\hat\delta_r^{(2)}\theta_{m}\rangle}{\langle\hat\delta_r^{(2)}\hat\delta_r^{(2)}\rangle}\ ,\\
&\theta_r^{(2)}(\boldsymbol{k})=T_2(k)\hat\delta_r^{(2)}(\boldsymbol{k})\ .
\end{aligned}
\end{eqnarray}
The $\mathcal{O}(2)$ reconstruction is $\theta_r=\theta_r^{(1)}+\theta_r^{(2)}$.

\section{Performance}
\label{sec:results}

\subsection{Lagrangian velocity}

\begin{figure}[ht!]
\plotone{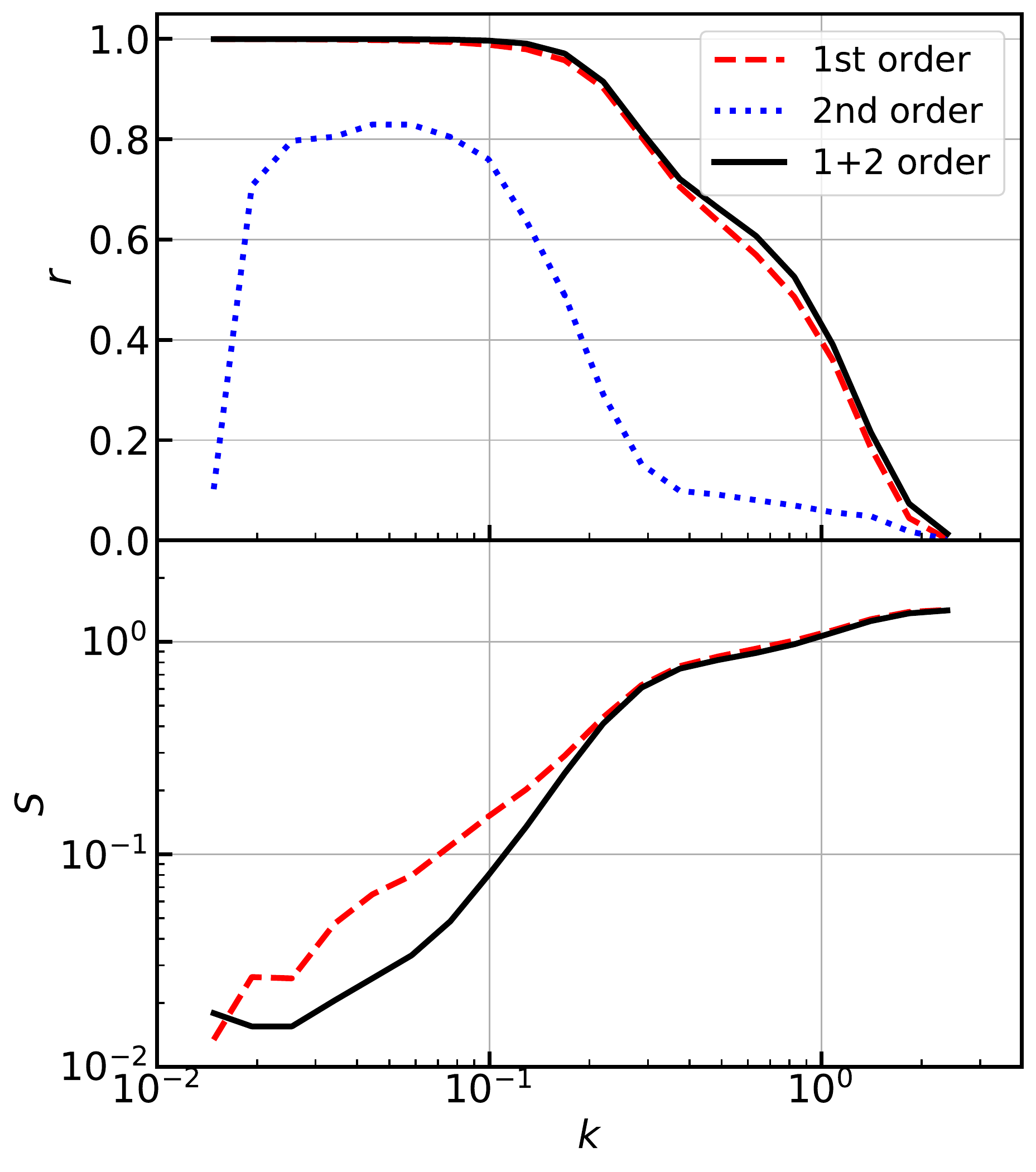}
\caption{
The upper panel shows the cross-correlation coefficient between $\delta^{(1)}_r$ and $\theta_t$,
$\delta^{(2)}_r$ and $\theta_m$, and $\delta_r=\delta^{(1)}_r+\delta^{(2)}_r$ and $\theta_t$.
They are presented as red dashed, blue dotted, and black solid lines, respectively.
The good cross-correlation is the basis for the velocity reconstruction in Lagrangian space.
The lower panel shows the stochasticity of the reconstructed velocity field relative to the true one in Lagrangian space.
It is defined as $S=\sqrt{2(1-r)}$ and amplifies the difference at large scales where $r$ is very close to unity.
\label{fig:lagxcc}}
\end{figure}

We first look at the reconstruction performance in Lagrangian space.
The red dashed line in the upper panel of Fig. \ref{fig:lagxcc} shows the cross-correlation coefficient between the reconstructed density field $\delta^{(1)}_r(\boldsymbol{q})$ and the true velocity divergence in Lagrangian space $\theta_t(\boldsymbol{q})$.
We found that at a large scale of $k<0.2\ \hmpc$, the cross-correlation is close to one,
and toward small scales the coefficient decreases.
At large scales, both the velocity and the displacement are linear.
This naturally leads to an almost perfect correlation at large scales.
At small scales, both the velocity and the displacement suffer from the nonlinear effects.
We expect the influence to be more severe in the velocity field.
These nonlinear effects change the small-scale velocity substantially and cause the loss of the correlation with the displacement.

We also plot the cross-correlation between the second-order term $\delta_r^{(2)}(\boldsymbol{q})$ and the residual velocity divergence $\theta_m(\boldsymbol{q})=\theta_t(\boldsymbol{q})-\theta_r^{(1)}(\boldsymbol{q})$ as a blue dotted line.
We also find a significant correlation, $r\sim 0.8$ at $0.02\ \hmpc<k<0.2\ \hmpc$,
implying that $\mathcal{O}(2)$ reconstruction could help.
However, how much adding this second-order term improves the result also depends on the power relative to the first-order term.
The result for the $\mathcal{O}(2)$ reconstruction is shown as a black solid line.
A Slight improvement is seen at $k<0.15\ \hmpc$ and $k> 0.4\ \hmpc$.

In the bottom panel we plot the stochasticity, defined as $S=\sqrt{2(1-r)}$.
This statistics amplifies the difference when $r\sim 1$.
The $\mathcal{O}(1)$ reconstruction result is plotted as a red dashed line and the $\mathcal{O}(2)$ reconstruction is a black solid line.
From this panel we see that $\mathcal{O}(2)$ reconstruction has a lower stochasticity than the standard method by a factor of $\sim 3$ at $0.02\ \hmpc<k<0.1\ \hmpc$.
This suppression of the stochasticity at large-scales is important for measuring the large scale effects such as the primordial non-Gaussianity by the sample variance cancellation technique (\cite{Munchmeyer18}).

\subsection{Eulerian velocity}

\begin{figure}[ht!]
\plotone{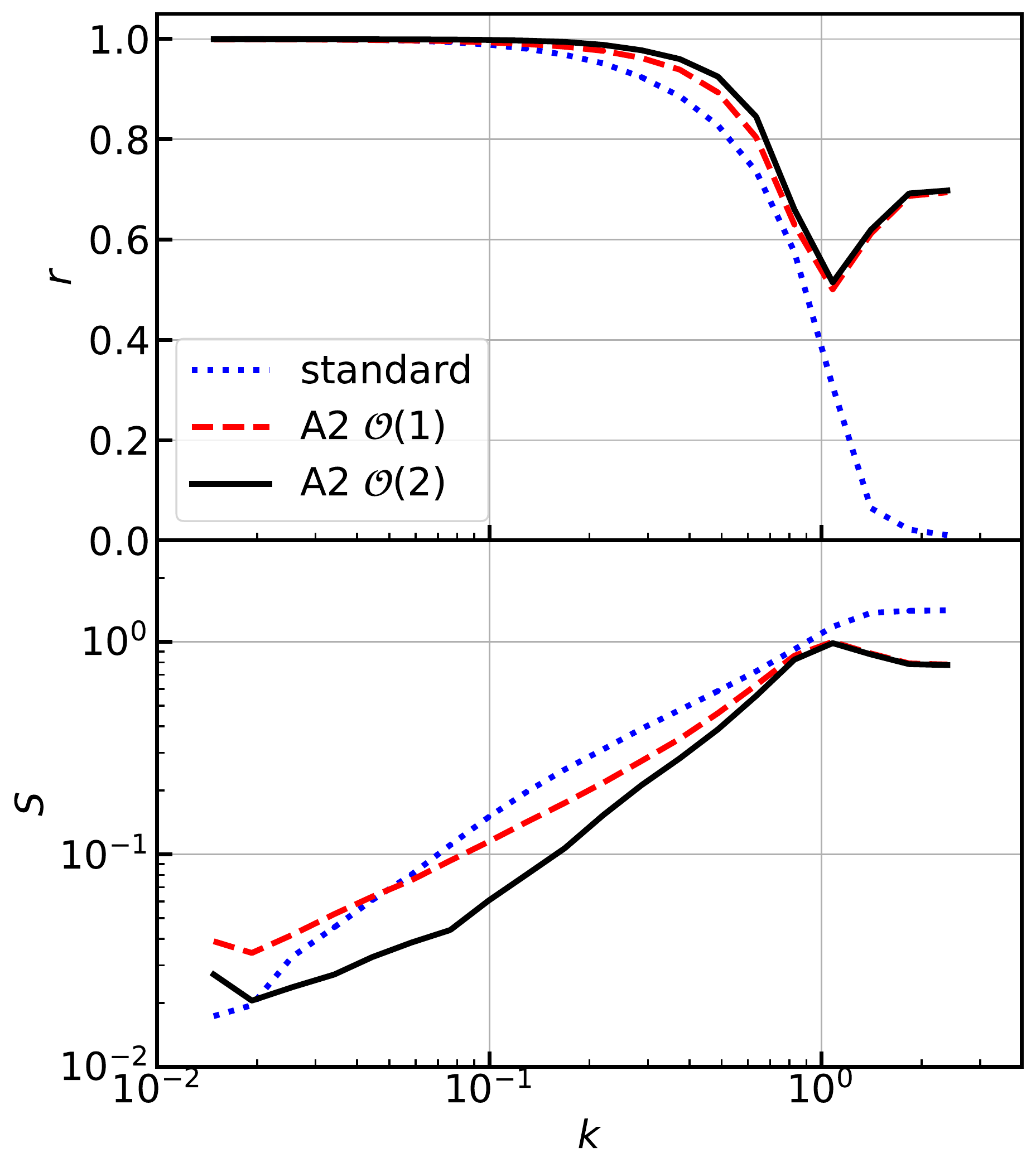}
\caption{The upper panel shows the cross-correlation coefficient between the true Eulerian velocity field and the reconstructed velocity field by the standard reconstruction (blue dotted line), $\mathcal{O}(1)$ reconstruction (red dashed line), and $\mathcal{O}(2)$ reconstruction (black solid line).
The lower panel shows the stochasticity induced by the reconstruction methods.
\label{fig:eulxcc}}
\end{figure}

After mapping the reconstructed Lagrangian velocity to Eulerian space by the displacement, we use the NP velocity assignment scheme to obtain the reconstructed velocity field.
Note that the true velocity field is obtained by the same NP assignment.
The cross-correlation coefficients between the two are presented in Fig. \ref{fig:eulxcc}.
The red dashed line, black solid line, and blue dotted line represent the results for $\mathcal{O}(1)$ reconstruction, $\mathcal{O}(2)$ reconstruction, and the standard reconstruction, respectively.
We find that the proposed $\mathcal{O}(1)$ reconstruction method performs better than the standard one at scales $k>0.1\ \hmpc$,
and the $\mathcal{O}(2)$ reconstruction further slightly improves the cross-correlation coefficient at scales $0.1\ \hmpc<k<1\ \hmpc$.

In the bottom panel of Fig. \ref{fig:eulxcc} we show the stochasticity for the above reconstruction methods.
Adding $\mathcal{O}(2)$ reconstruction suppresses the stochasticity at scales $k<1\ \hmpc$ relative to the $\mathcal{O}(1)$ reconstruction,
and it performs better than the standard reconstruction method at scales $0.02\ \hmpc<k<1\ \hmpc$.

One obvious feature is that the cross-correlation coefficient decreases toward small scales for $k<1\ \hmpc$ but increases at $k>1\ \hmpc$.
However, the result at $k>1\ \hmpc$ is suspicious due to the fact that this scale is close to the Nyquist frequency of the analysis and this good correlation between the reconstructed and the true velocity may partially come from the same systematics by the same velocity assignment.

\begin{figure}[ht!]
\plotone{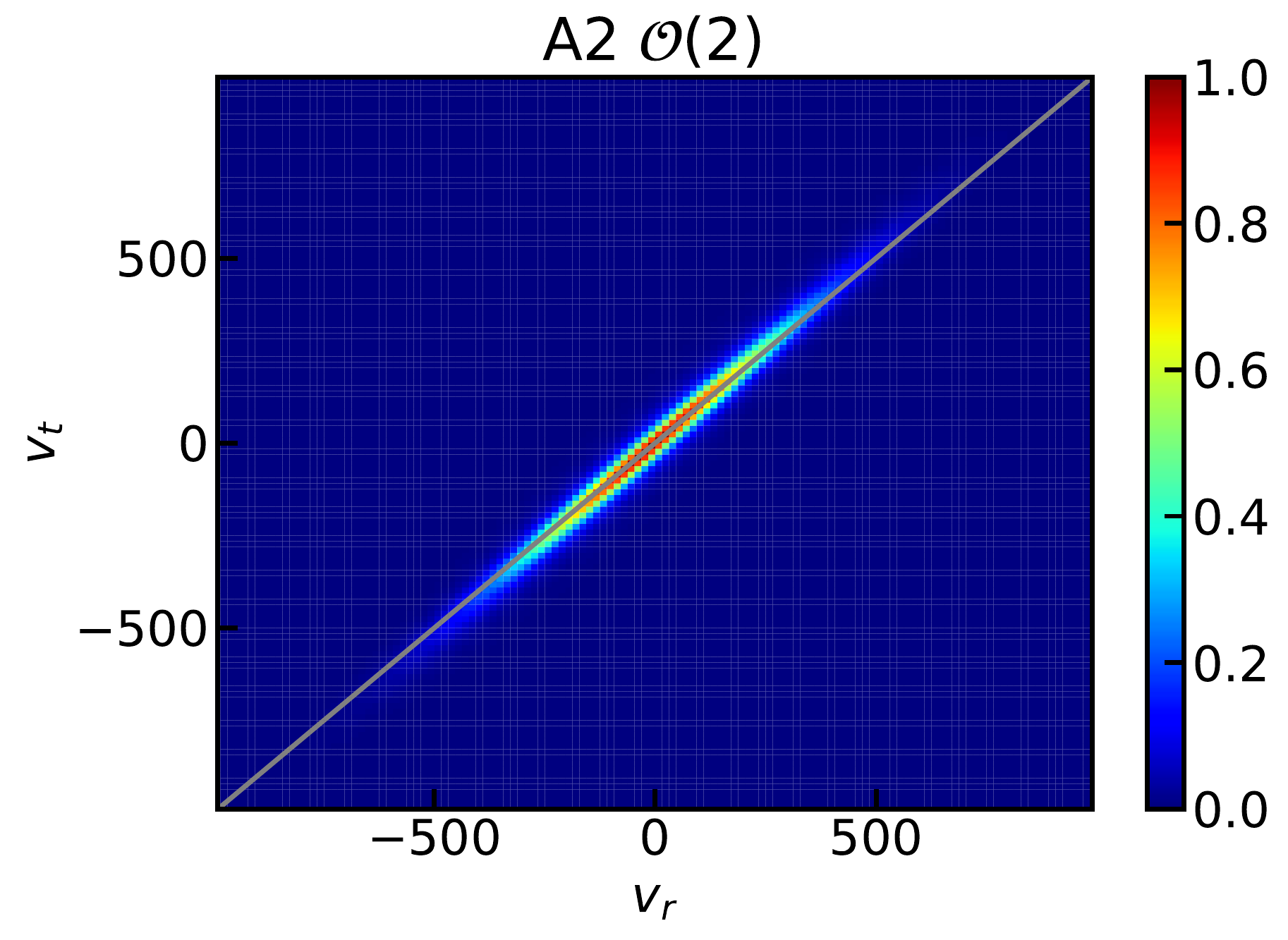}\\
\plotone{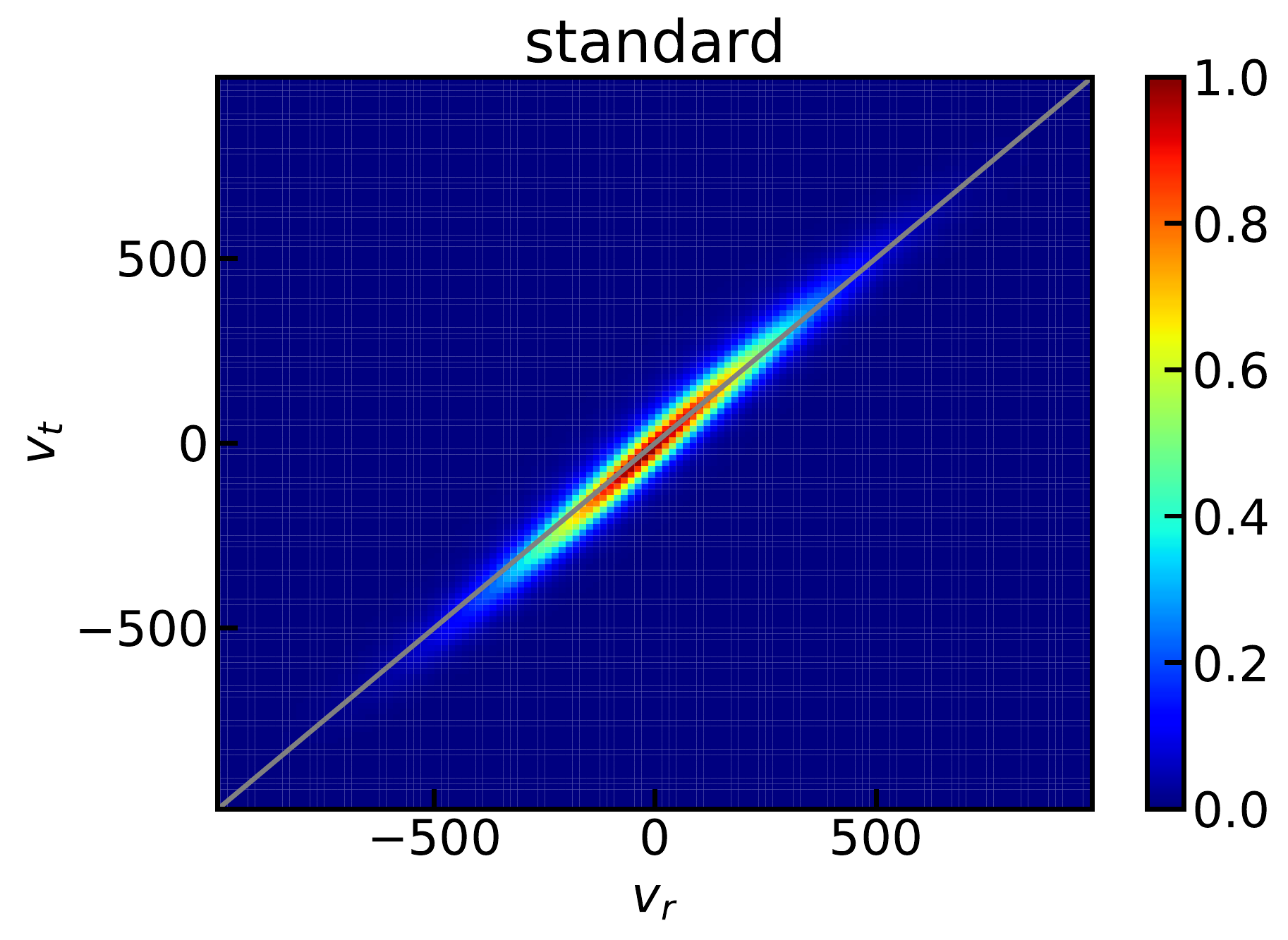}
\caption{The upper panel shows the 2D histogram for the true velocity and the reconstructed velocity by the $\mathcal{O}(2)$ reconstruction.
The lower panel is for the true velocity and the velocity reconstructed by the standard method.
The new reconstruction results in a tighter relation between the reconstructed velocity and the true one.
\label{fig:eulscatter}}
\end{figure}

We compare the reconstructed velocity with the true one at each grid point in Fig. \ref{fig:eulscatter},
which shows the two-dimensional histogram, with the horizontal axis being the reconstructed velocity while the vertical axis is the true velocity.
The color indicates the relative counts normalized to unity.
The upper panel shows the result from $\mathcal{O}(2)$ reconstruction and the lower panel shows the result from the standard method.
Compared to the standard reconstruction, we observe an obvious slimmer contour for the $\mathcal{O}(2)$ reconstruction.

\begin{figure}[ht!]
\plotone{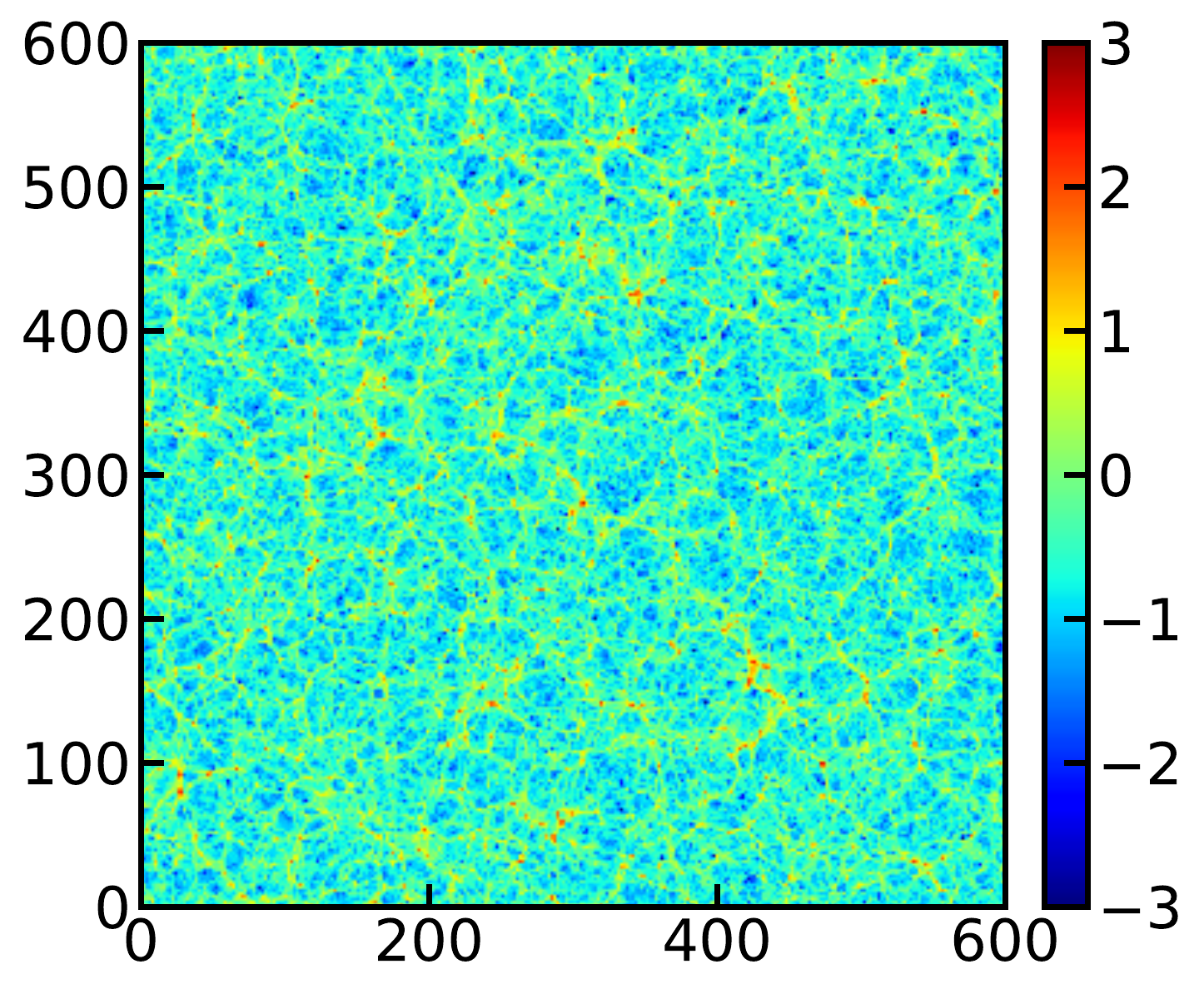}\\
\hspace{1mm}\plotone{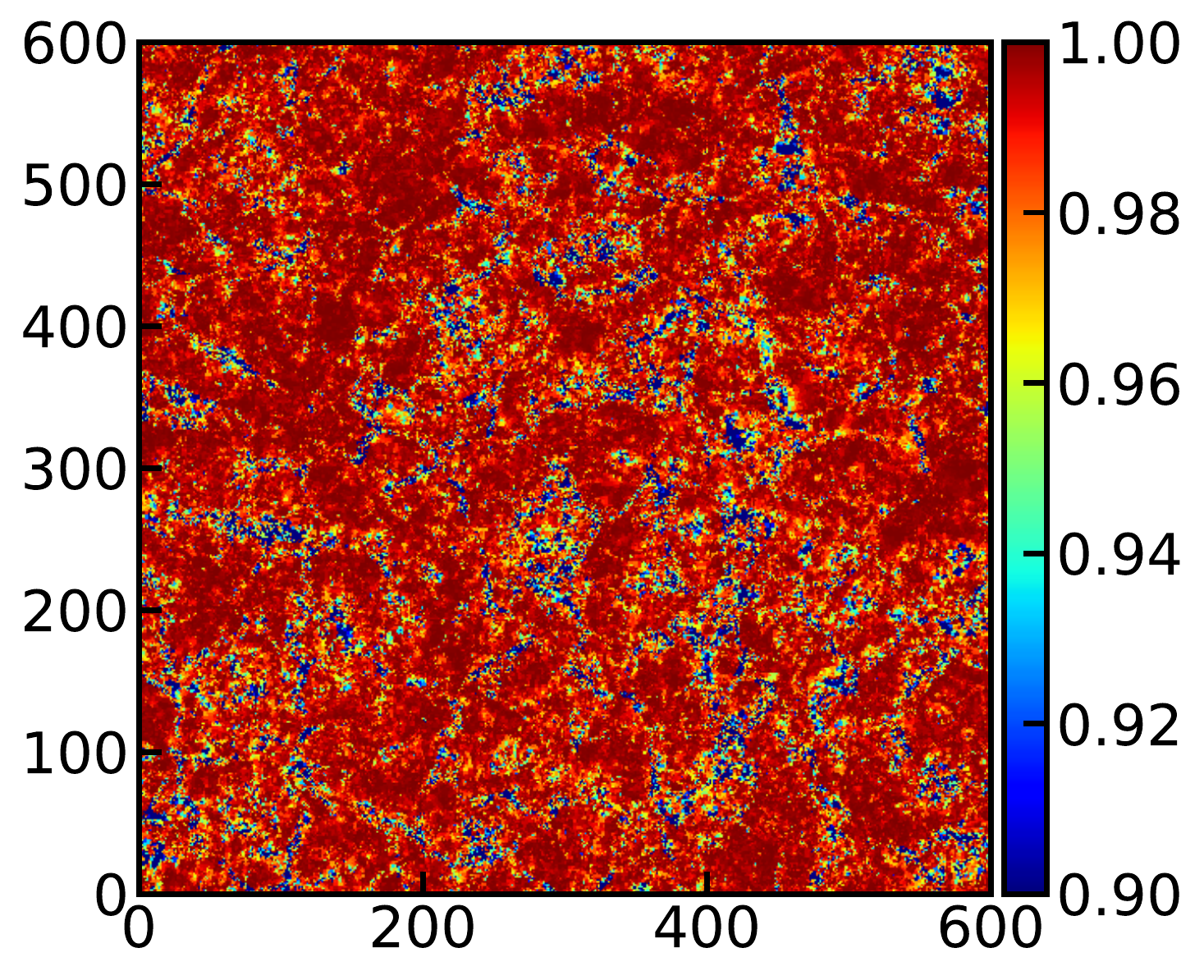}\\
\hspace{1mm}\plotone{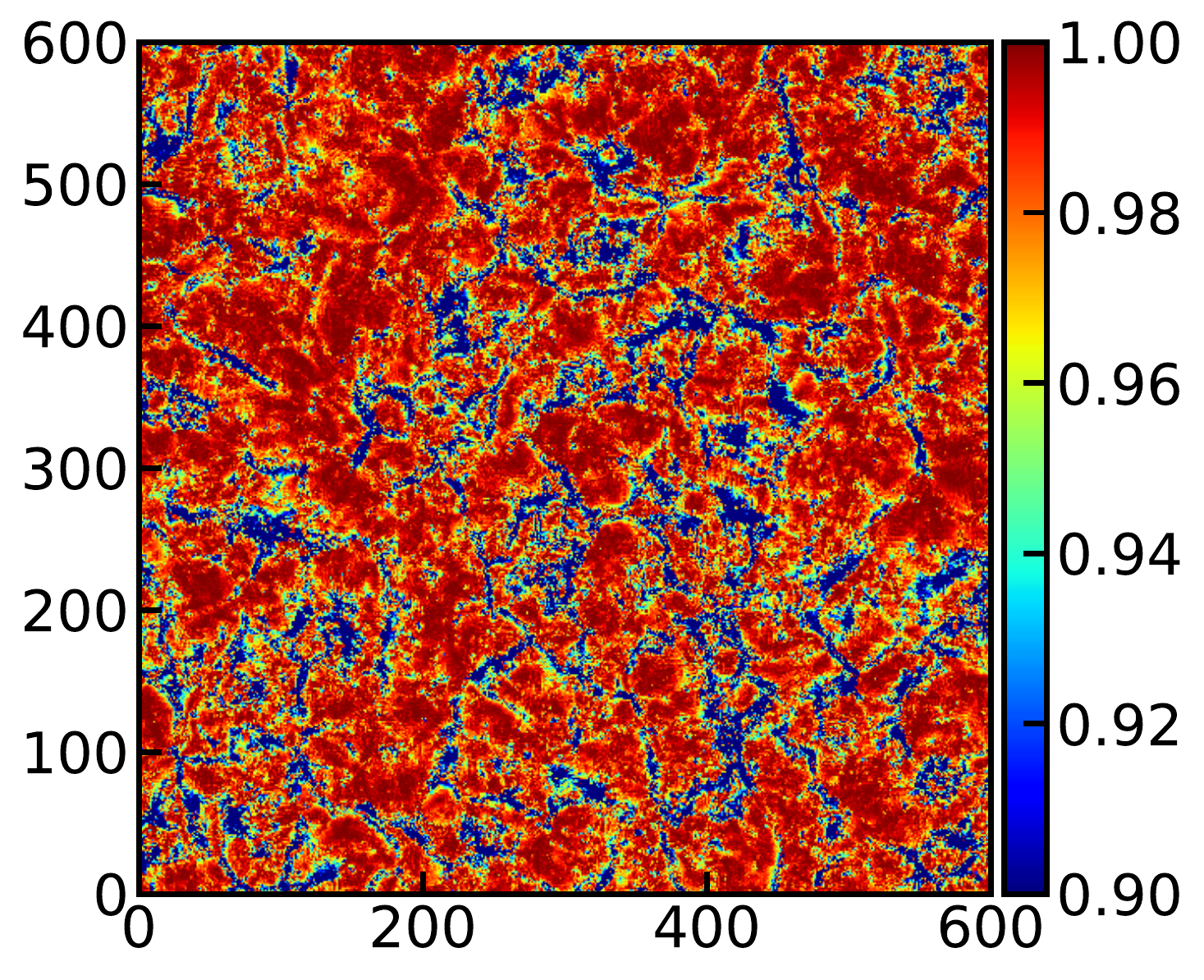}
\caption{
Top: a slice of DM density field $\log(1+\delta)$.
Middle: a slice of the cosine angle between $v_t$ and $v_r$ reconstructed by the $\mathcal{O}(2)$ reconstruction.
Bottom: a slice of the cosine angle between $v_t$ and $v_r$ reconstructed by the standard method.
Large misalignment appears in the high-density region.
\label{fig:cosanglemap}}
\end{figure}

To quantify the performance, we check the direction and the amplitude of the reconstructed velocity.
We define the cosine angle between it and the true one as
\begin{equation}
\mu=\frac{\boldsymbol{v}_t\cdot\boldsymbol{v}_r}{|\boldsymbol{v}_t||\boldsymbol{v}_r|}\ .
\end{equation}
We plot this cosine angle for the $\mathcal{O}(2)$ reconstruction and the standard one of one slice in the middle and bottom panels of Fig. \ref{fig:cosanglemap}, respectively.
Also plotted is the dark matter density of the same slice in the top panel.
Both reconstruction methods perform worse in the high-density region, i.e. the highly nonlinear region.
We find that for the $\mathcal{O}(2)$ reconstruction, the region with $\mu<0.95$ (green to blue color) occupies far less volume than the standard reconstruction result,
indicating that the reconstruction performs well down to the nonlinear region.

\begin{figure}[ht!]
\plotone{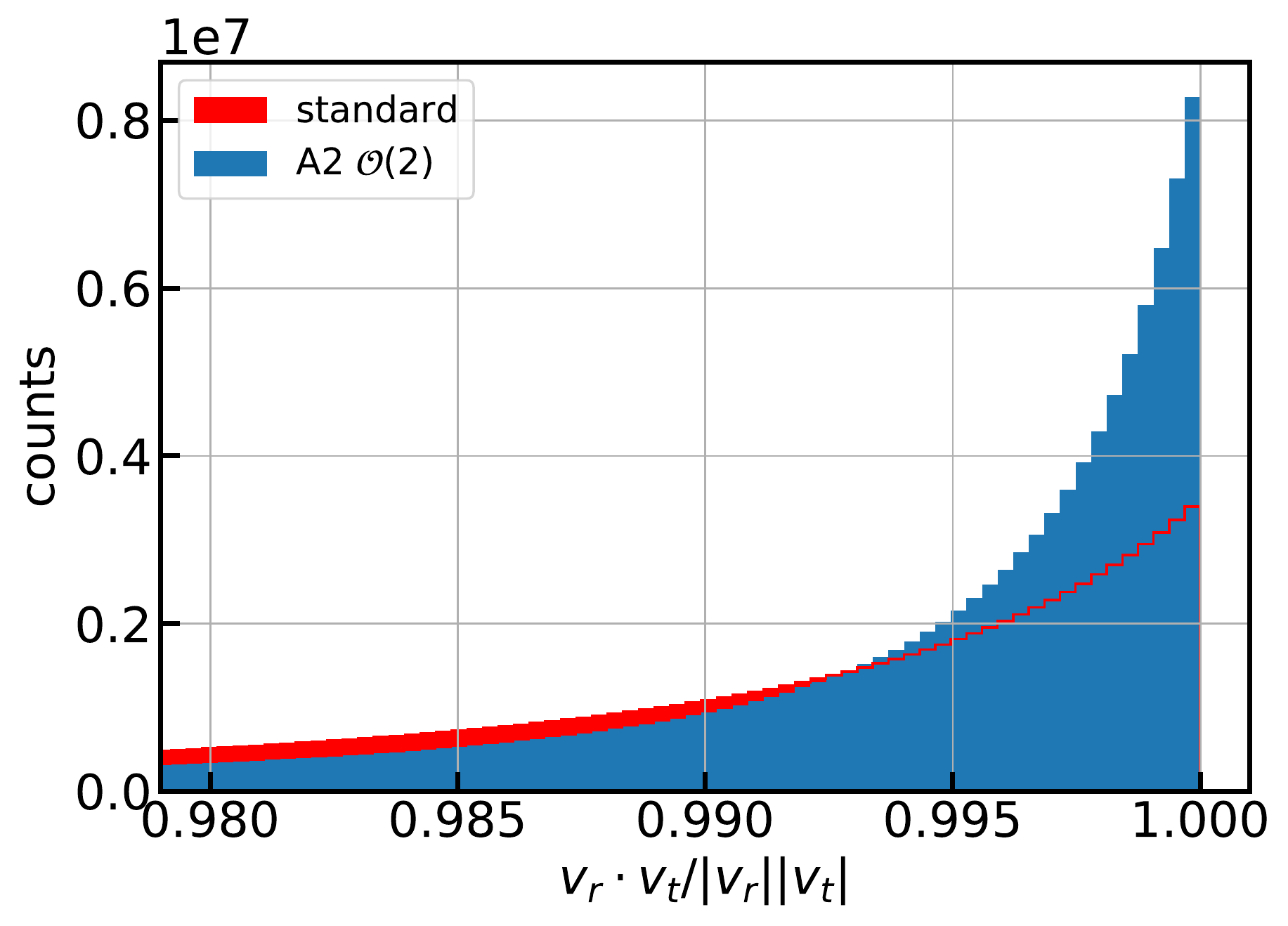}
\caption{
Distribution of the cosine angle between $\boldsymbol{v}_t$ and $\boldsymbol{v}_r$.
The red histogram is for standard reconstruction, while the blue one is for the $\mathcal{O}(2)$ reconstruction.
The $\mathcal{O}(2)$ reconstruction has much more pixels with $\mu>0.993$.
The average cosine angle is $0.9577$ for standard reconstruction, corresponding to misalignment of $16^{\circ}.72$.
For nonlinear reconstruction it is $0.9790$ and $11^{\circ}.77$.
\label{fig:cosangledist}}
\end{figure}

The mean of the $\mu$ for the A2 $\mathcal{O}(2)$ reconstruction is $0.977$, while the standard method has the mean $\langle\mu\rangle=0.958$.
This corresponds to the mean misalignment angle of $12.31 \deg$ and $16.66 \deg$ for the A2 $\mathcal{O}(2)$ reconstruction and the standard one, respectively.
We also plot the histogram for the cosine angle $\mu$ in Fig. \ref{fig:cosangledist}.
The $\mathcal{O}(2)$ reconstruction (blue histogram) has much more pixels with very good direction reconstruction ($\mu>0.995$) than the standard method (red histogram).

\begin{figure}[ht!]
\plotone{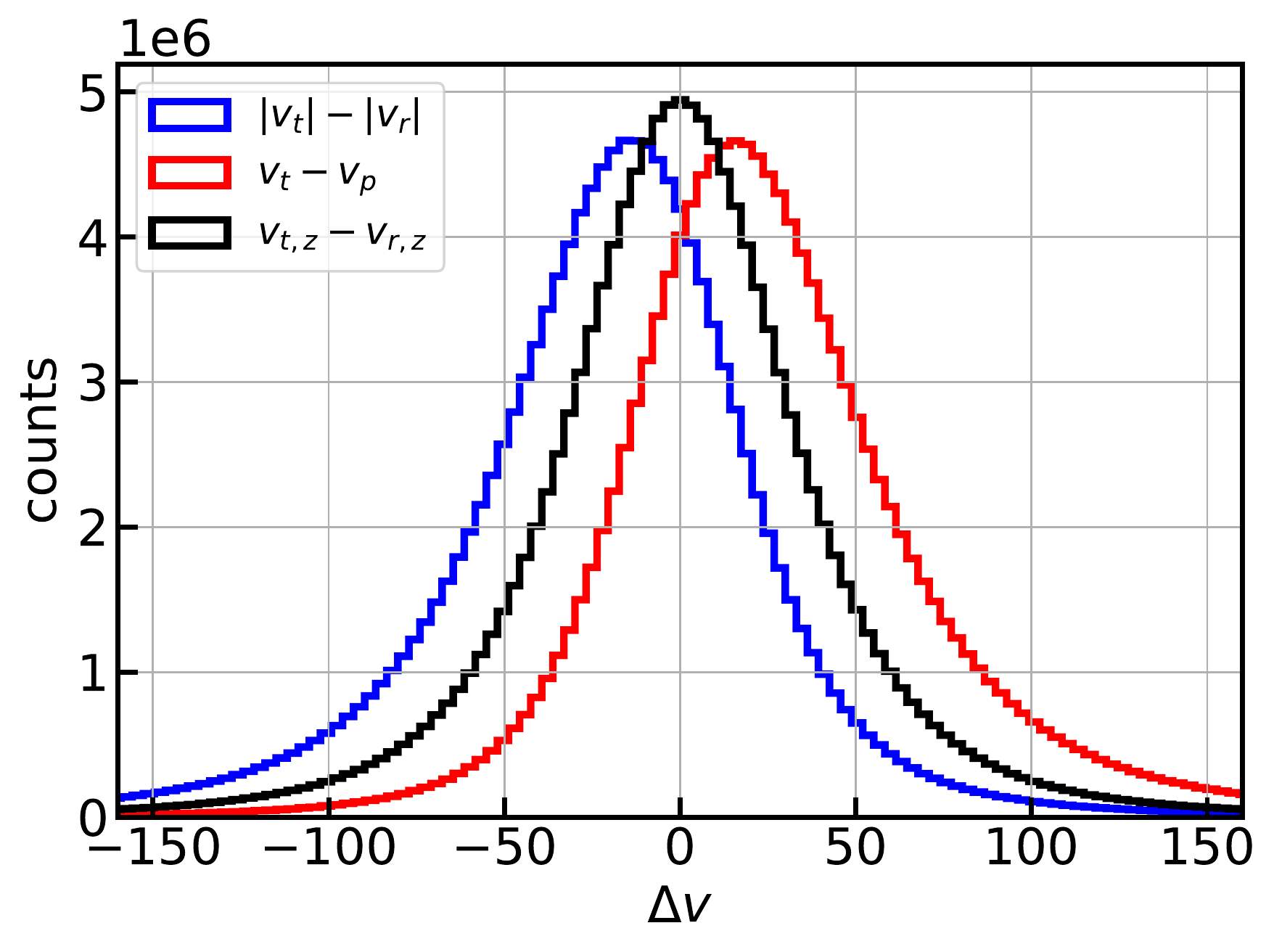}\\
\plotone{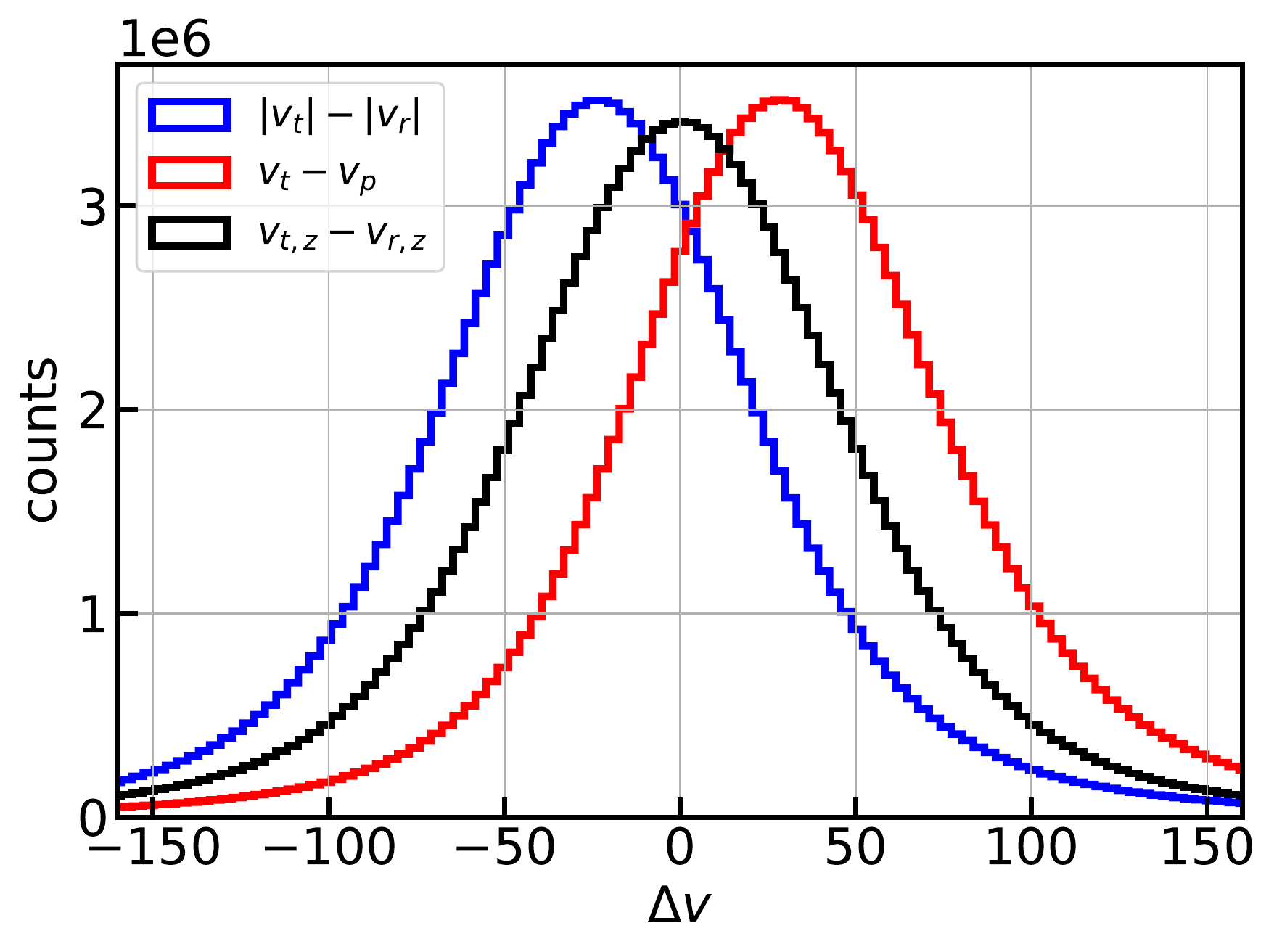}
\caption{Blue: the magnitude difference between $\boldsymbol{v}_t$ and $\boldsymbol{v}_r$.  
Red: the difference between $\boldsymbol{v}_t$ and $\boldsymbol{v}_r$ projected on $\boldsymbol{v}_t$. 
Black: the difference between $\boldsymbol{v}_t$ and $\boldsymbol{v}_r$ in $z$-direction.
The top panel is the result of the A2 $\mathcal{O}(2)$ reconstruction method, 
and the bottom panel is the result of the standard reconstruction.
The $\mathcal{O}(2)$ reconstruction result presents a smaller amplitude bias and smaller scatter in the distribution.
\label{fig:diffdist}}
\end{figure}

We also check whether the amplitude of the velocity is reconstructed well.
Here we define three kinds of velocity amplitude difference.
The first one is the difference between the true velocity amplitude and the reconstructed one.
The second is the difference between the true velocity and the projection of the reconstructed velocity on the true one, i.e. $\boldsymbol{v}_{p}=\boldsymbol{v}_r\cdot\boldsymbol{v}_t/|v_t|$.
The last one is the difference between the velocity component in the $z$-direction.

The results are shown in Fig. \ref{fig:diffdist}, in which the distribution of the velocity amplitude difference is plotted.
The top plot is the result from the $\mathcal{O}(2)$ reconstruction, while the bottom plot is from the standard one.
For the first and second distributions (blue and red histogram), the $\mathcal{O}(2)$ reconstruction result has a peak closer to zero than the standard reconstruction, i.e. a smaller reconstruction bias in the amplitude.
Furthermore, the $\mathcal{O}(2)$ reconstruction also has smaller scatters in these distributions than the standard one.
For the last statistics, the velocity difference in one direction, the standard method is expected to produce a mean of zero by design.
For the $\mathcal{O}(2)$ reconstruction, we also find this statistics has a mean of zero,
and the width of the distribution is much narrower than the standard one.
All the above statistics show that the $\mathcal{O}(2)$ reconstruction recovers the velocity amplitude relatively better than the other reconstruction methods.

\section{Ramifications}
\label{sec:remi}

\subsection{Using simulated displacement}

\begin{figure}[ht!]
\plotone{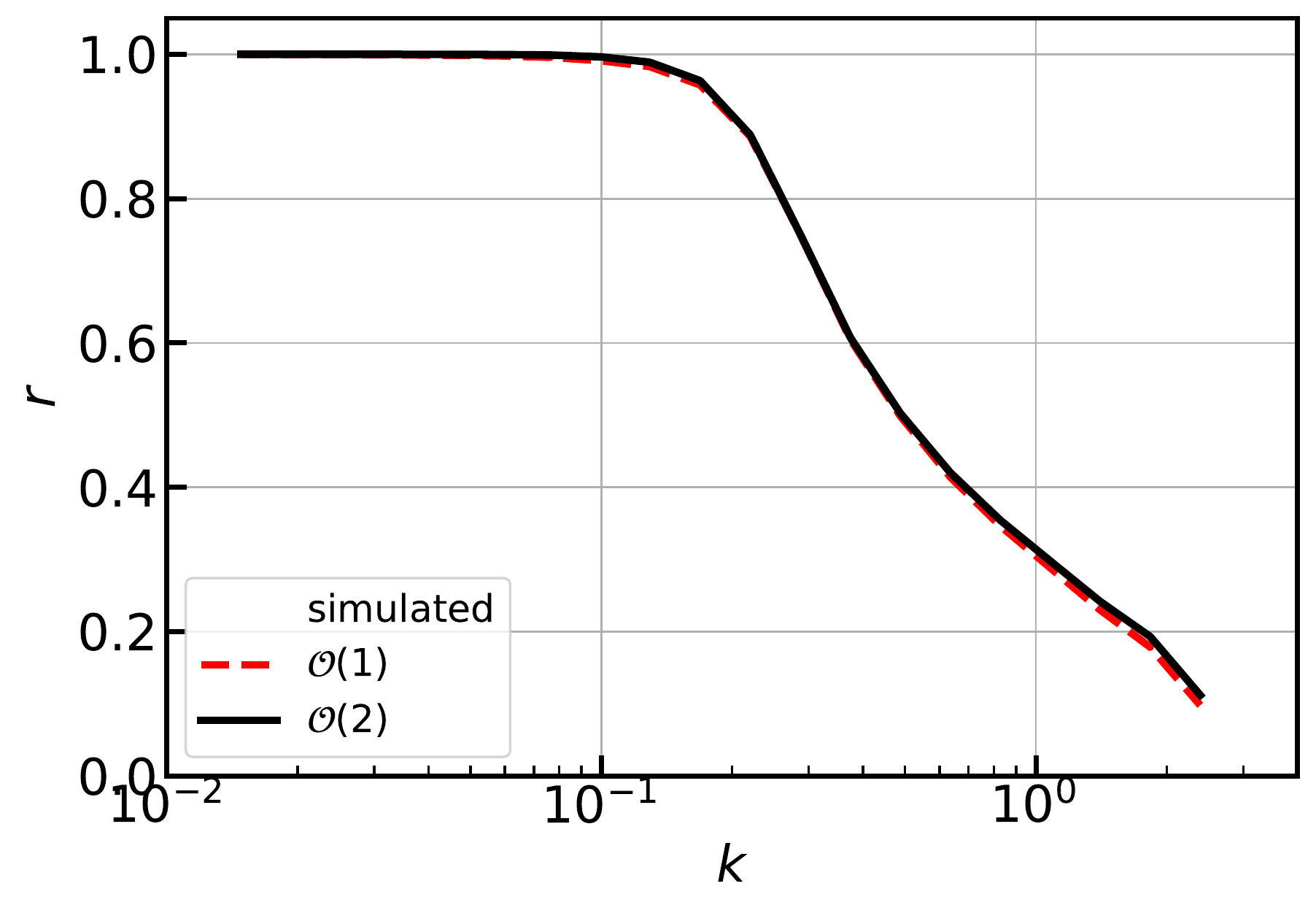}\\
\plotone{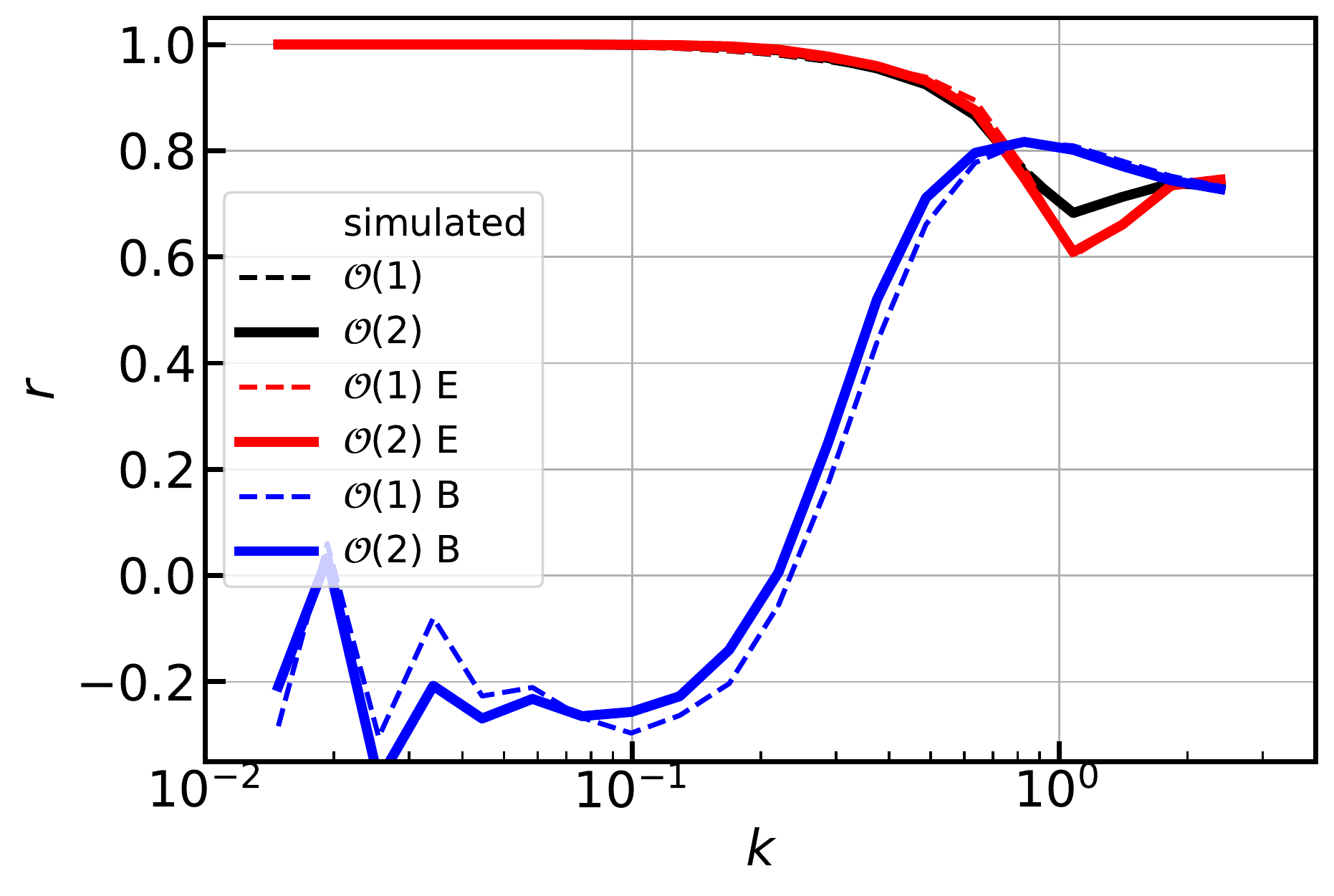}
\caption{The result of the reconstruction using the simulated displacement.
The top panel is the cross-correlation coefficient between $\boldsymbol{v}_r(\boldsymbol{q})$ and $\boldsymbol{v}_t(\boldsymbol{q})$ in Lagrangian space (red dashed line for $\mathcal{O}(1)$ and black solid line for $\mathcal{O}(2)$ reconstruction).
The bottom panel is the cross-correlation coefficient between $\boldsymbol{v}_r(\boldsymbol{x})$ and $\boldsymbol{v}_t(\boldsymbol{x})$ in Eulerian space.
The total, E-mode, and B-mode component results are presented in black, red, and blue lines.
\label{fig:simurecon}}
\end{figure}

The above results slightly depend on the performance of the detailed reconstruction algorithm.
We compare the performance of the three different algorithms in the Appendix.
Here we want to know the upper limit of the velocity reconstruction in Lagrangian space.
Thus, in this subsection we present the result under the assumption that the displacement estimation is perfect, i.e. the reconstruction is performed using the simulated displacement instead of the reconstructed one.
The cross-correlation coefficient in Lagrangian space is shown in the top panel of Fig. \ref{fig:simurecon}.
The red dashed line is the result from $\mathcal{O}(1)$ reconstruction and the black solid line is for $\mathcal{O}(2)$ reconstruction.
In this case we see that the improvement from including the second-order term is very mild.
The nonlinear displacement already includes almost all the information that can be used to reconstruct the nonlinear velocity field.

In the bottom panel we show the cross-correlation in Eulerian space.
The performance is better than the results using reconstructed displacement.
The cross-correlation coefficient between the reconstructed velocity and the true one reach $0.7$ at $k\sim 1\ \hmpc$ (black solid line).
No obvious difference between $\mathcal{O}(1)$ and $\mathcal{O}(2)$ reconstruction is observed and the black solid and black dashed lines overlap with each other.
We further decompose both the true and the reconstructed velocity field into curl-free E-mode and divergence-free B-mode and correlate them separately.
The cross-correlation for the E-mode is slightly lower than the total velocity at scale $k\sim 1\ \hmpc$.
The B-mode cross-correlation coefficient is $\sim -0.2$ at $k<0.1\ \hmpc$ and rapidly increases to $\sim 0.8$ at $k\sim0.7\ \hmpc$.
Note that the observed B-mode suffers from severe systematics in this configuration (refer to Fig. 12 in \cite{Zheng13}) and the power spectrum changes significantly as the simulation configuration changes.
We predict that the observed cross-correlation for the B-mode between the true velocity and the reconstructed one mainly comes from the systematics induced by the velocity assignment method, the finite volume effect, and the aliasing effect.

\subsection{Considering the B-mode in Lagrangian space}
\label{sec:bmode-in}

\begin{figure*}[ht!]
\plottwo{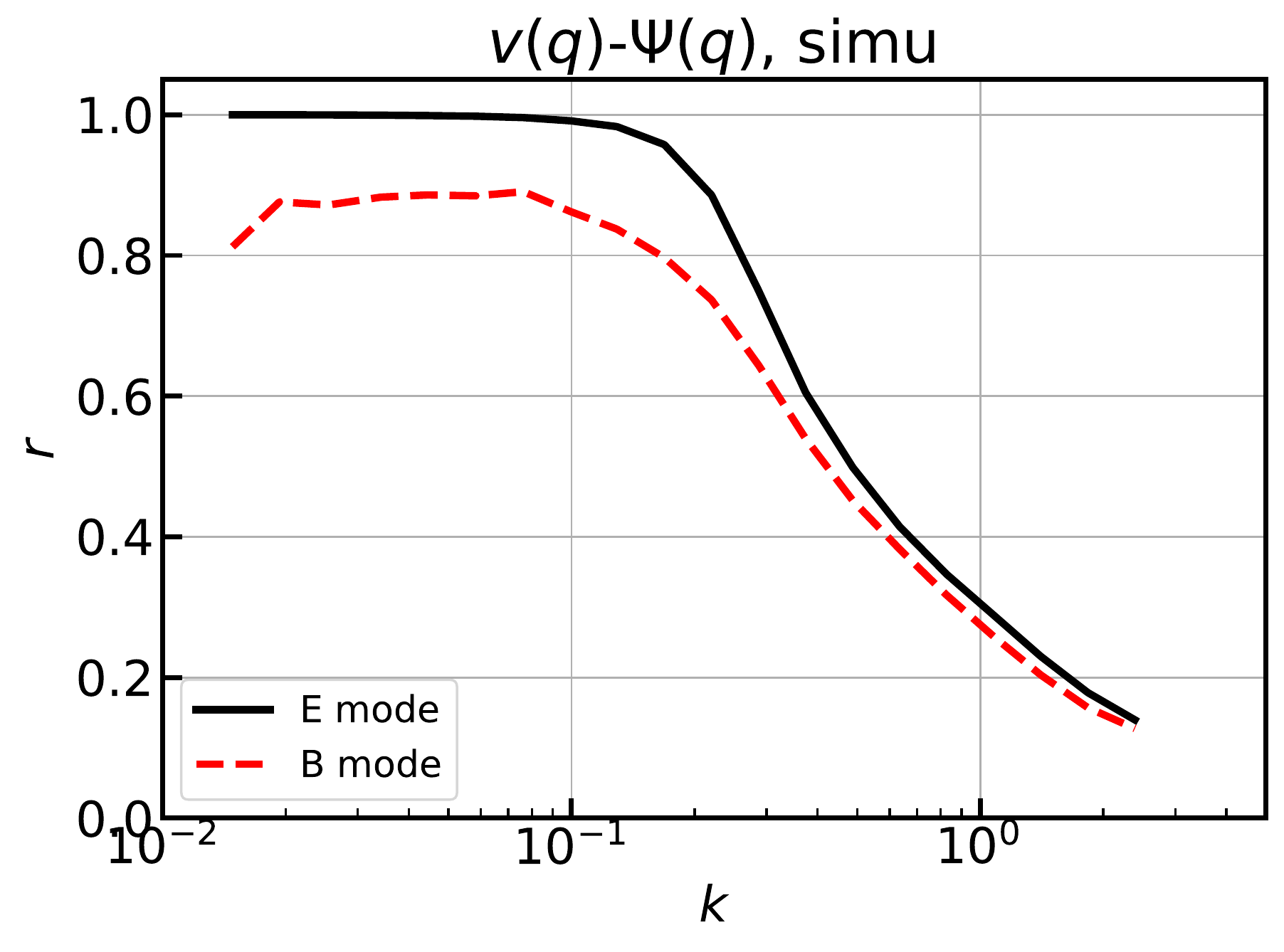}{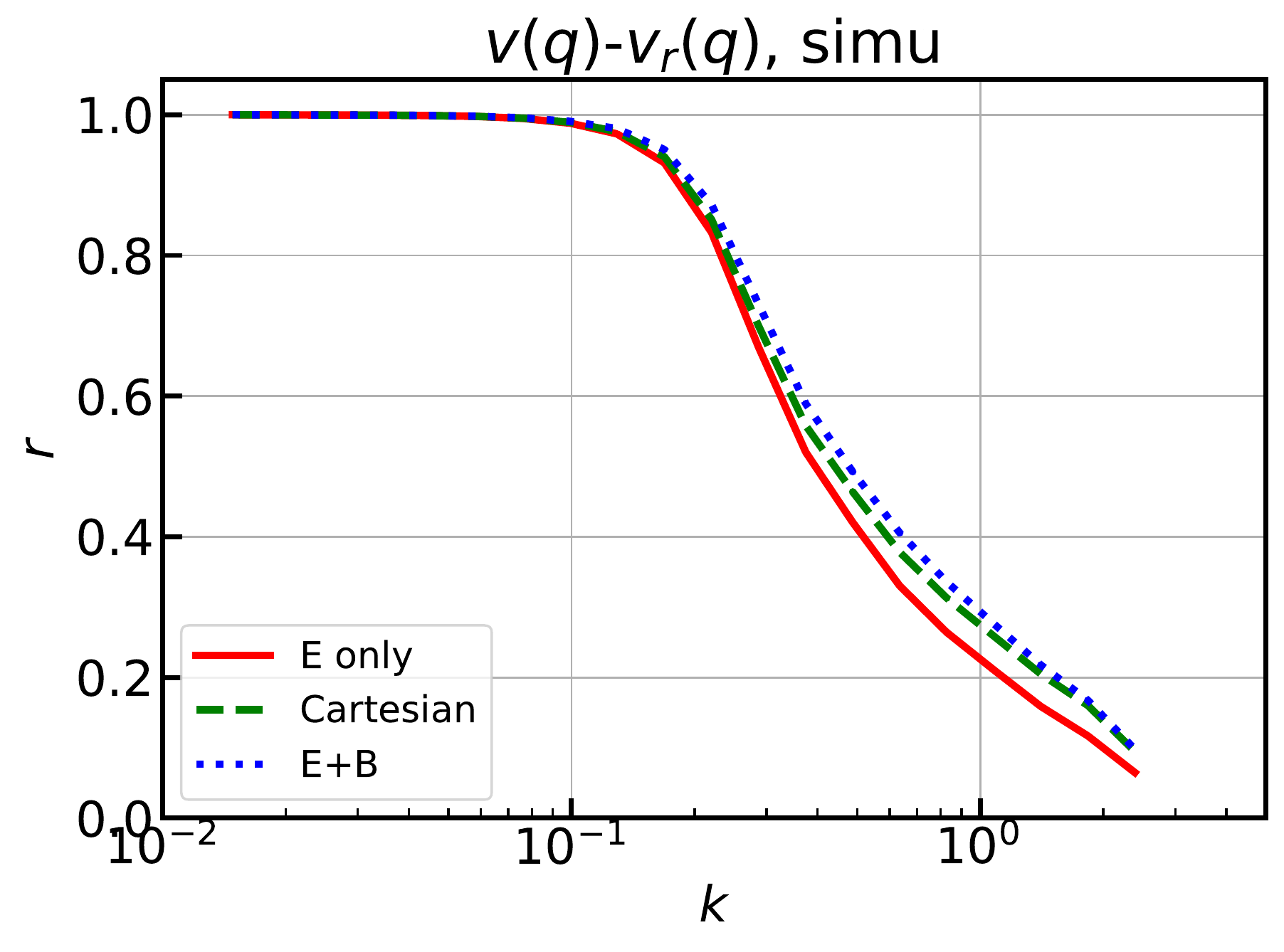}\\
\plottwo{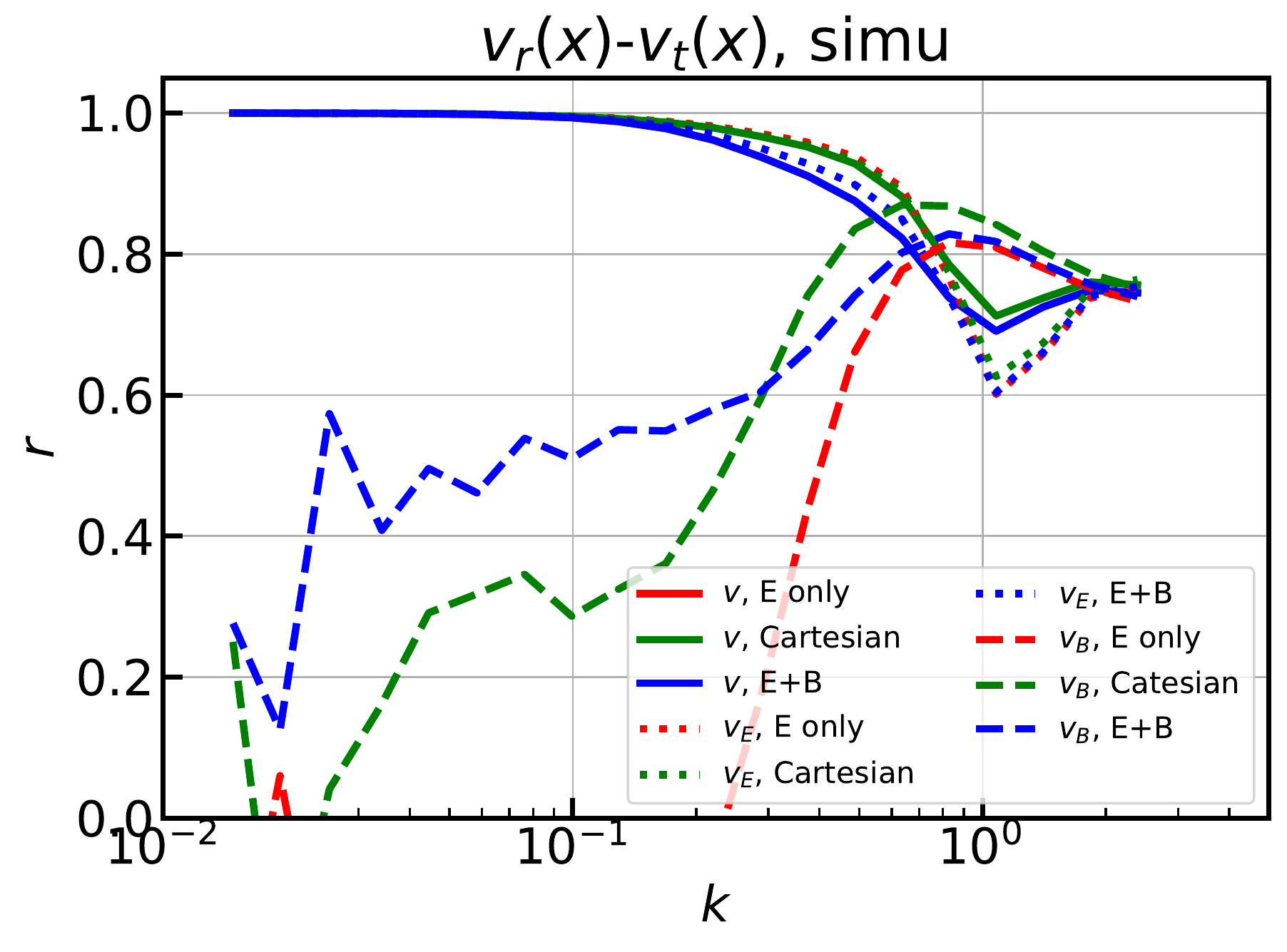}{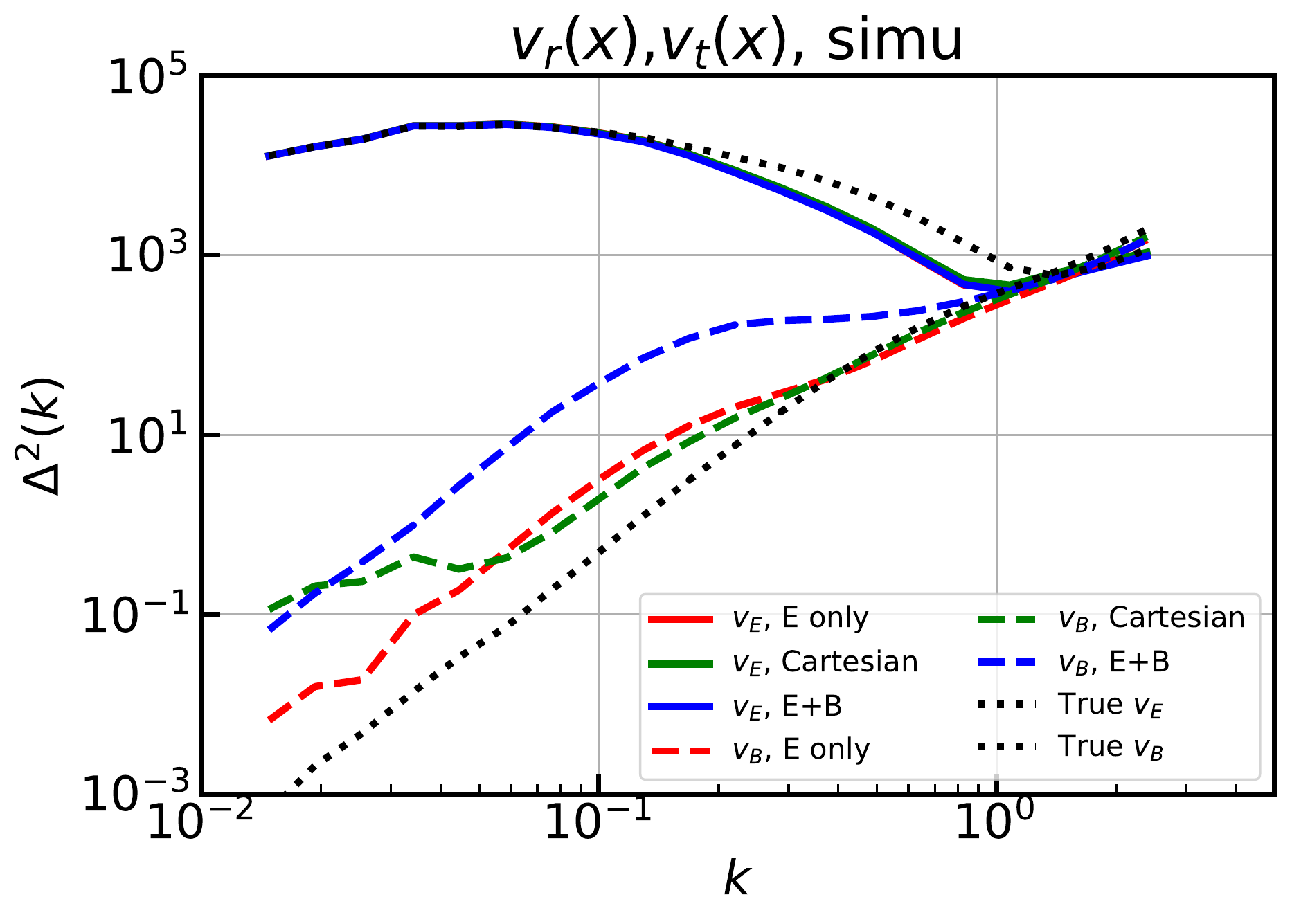}
\caption{The upper left panel shows the cross-correlation coefficient between the Lagrangian velocity and displacement in simulation.  The E-mode correlation is shown as a black solid line and the result for B-mode component is shown as a red dashed line.
The upper right panel shows the the cross-correlation coefficients between the reconstructed velocity and the true one in Lagrangian space.  The three cases correspond to E-mode-only reconstruction, reconstruction for three Cartesian directions separately, and reconstruction for E- and B-mode separately.
The bottom left panel is the cross-correlation between the reconstructed velocity and the true one in Eulerian space for different cases.
The bottom right panel is the results for power spectrum comparison.
\label{fig:bmode-in}}
\end{figure*}

In the above subsection, the reconstructed B-mode in Eulerian space comes from the nonlinear mapping and is converted from the E-mode in Lagrangian space.
Due to the nonlinear evolution of the universe, both the velocity and the displacement have a B-mode component (see e.g. \cite{Chan14}) both in Eulerian and Lagrangian space.
Here we are curious about whether the Lagrangian space B-mode correlation helps in the reconstruction.
The cross-correlations between the two from simulations are shown as the red dashed line in the upper left panel of Fig. \ref{fig:bmode-in}.
Also plotted is the cross-correlation from the E-mode as a black solid line.
At large scales, the B-mode velocity and B-mode displacement also have a large cross-correlation coefficient, $\sim 0.9$.
This B-mode correlation could help in the velocity reconstruction.
This also implies that the Cartesian components of the velocity and displacement field contain extra correlations other than the divergence of the two. 
In the upper right panel we show the cross-correlation coefficient between the true velocity and the velocity reconstructed (1) using the transfer function measured only from the E-mode, ($\theta_r(\boldsymbol{k})=T^E(k)\delta_r(\boldsymbol{k})$): (2) using the transfer function measured from the Cartesian components, ($\boldsymbol{v}_{r,i}(\boldsymbol{k})=T_i(k)\boldsymbol{\Psi}_{r,i}(\boldsymbol{k})$, and $i$ runs for 3 Cartesian components): and (3) using the transfer functions for the E- and B-modes separately and summing the two reconstructed velocity fields,
($\theta_r(\boldsymbol{k})=T^E(k)\delta_r(\boldsymbol{k})$ and $\boldsymbol{v}^B_{r,i}(\boldsymbol{k})=T^B(k)\boldsymbol{\Psi}^B_{r,i}(\boldsymbol{k})$).

Adding the B-mode improves the velocity reconstruction in Lagrangian space.
Reconstruction from the Cartesian components also improves the reconstruction, but it performs worse than directly adding B-mode reconstruction since it neglects the correlation between different Cartesian components.

However, this improvement in Lagrangian space is mainly at scales $k>0.2\ \hmpc$ and is not mapped into Eulerian space.
In the bottom left panel of Fig. \ref{fig:bmode-in} we present the cross-correlation coefficient for the above three cases and for the total, E-mode, and B-mode components separately.
We find that including the B-mode information in Lagrangian space leads to worse performance in Eulerian space.
We argue that the noise in the B-mode displacement (the part not correlated with the velocity) is converted into E-mode velocity in Eulerian space,
thus contaminating the velocity reconstruction instead of improving it.
The bottom right panel shows the E- and B-mode power spectrum of the reconstructed velocity field and the true one.
Including the B-mode correlation in Lagrangian space does not change the E-mode power spectrum in Eulerian space much,
but the Eulerian B-mode power spectrum is changed significantly.
This also implies that the B-mode is mainly a noise.

\begin{figure*}[ht!]
\plottwo{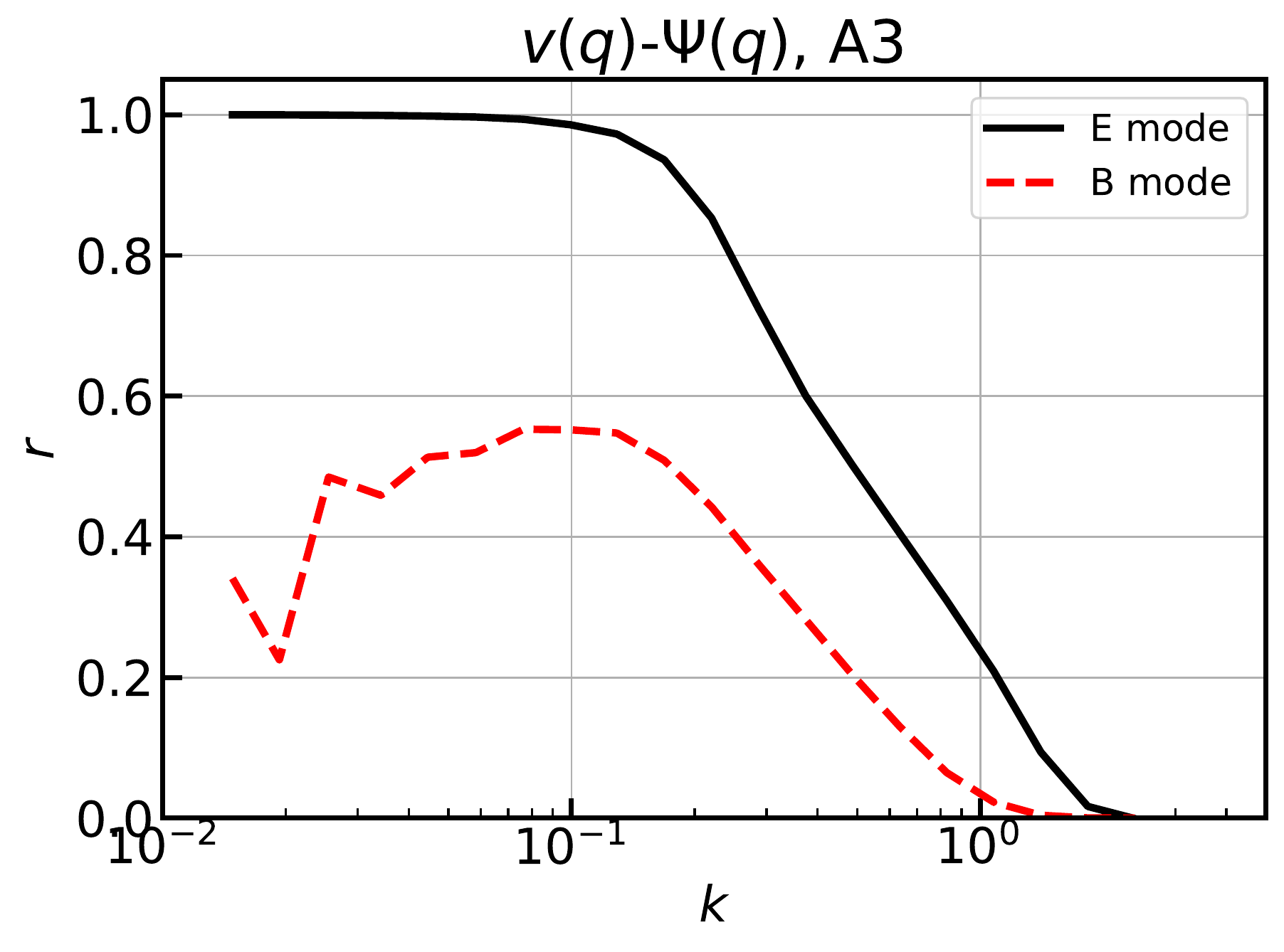}{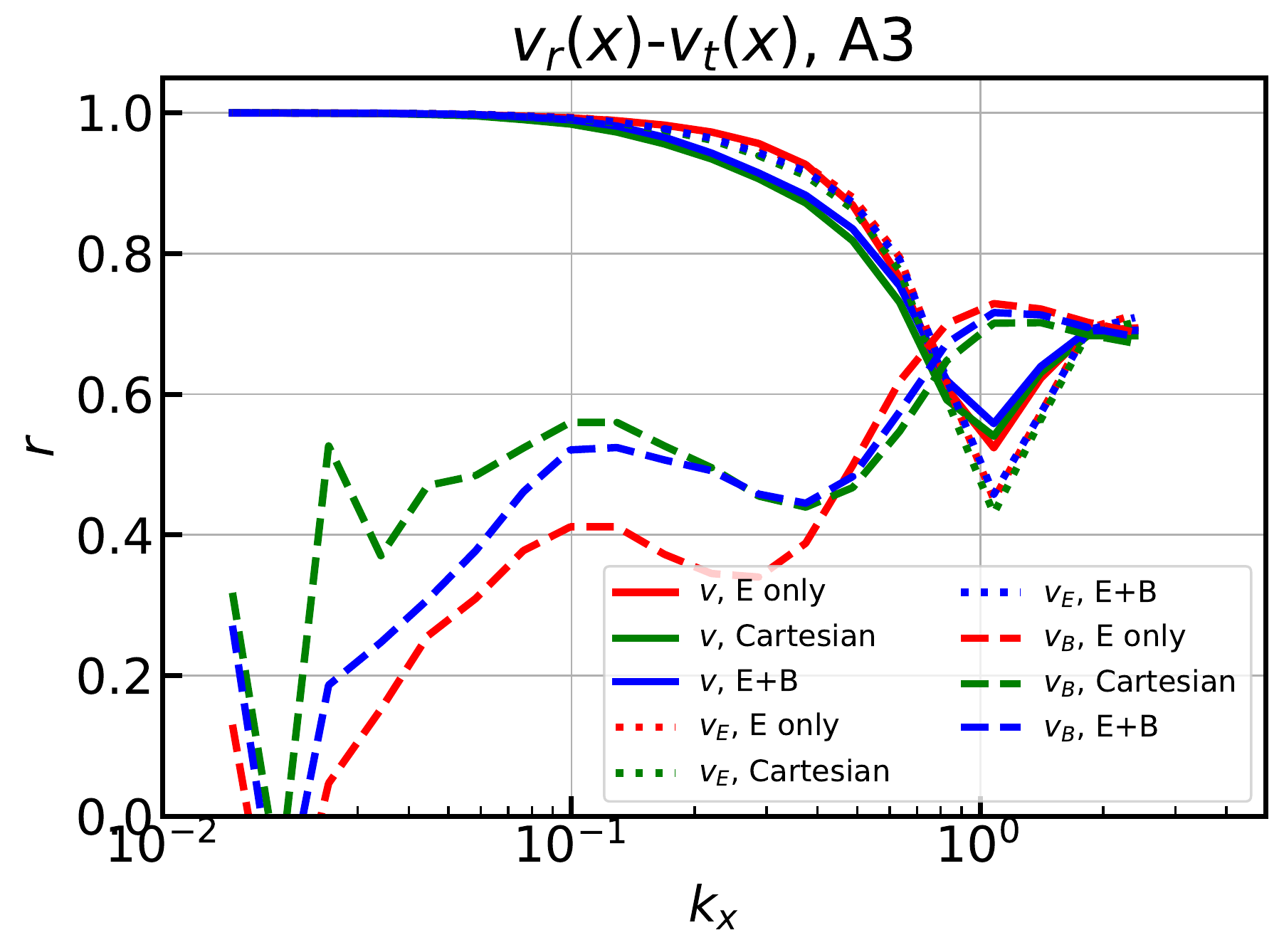}\\
\caption{Same as the corresponding panels in Fig. \ref{fig:bmode-in}, but for the reconstructed displacement by the A3 method.
\label{fig:bmode-in2}}
\end{figure*}

The A1 and A2 reconstruction algorithm has no B-mode displacement by design, but the A3 reconstruction algorithm does.
In the left panel of Fig. \ref{fig:bmode-in2}, we find that
this estimated B-mode displacement has a weaker correlation ($r\sim 0.5$) with the true B-mode velocity compared with the case using the real simulated displacement.
Thus, it suffers from more severe noise than the previous case.
It is expected that the reconstruction by adding the B-mode in Lagrangian space performs worse in Eulerian space.
This is observed in the right panel of Fig. \ref{fig:bmode-in2}).

\subsection{Using linear displacement}

\begin{figure}[ht!]
\plotone{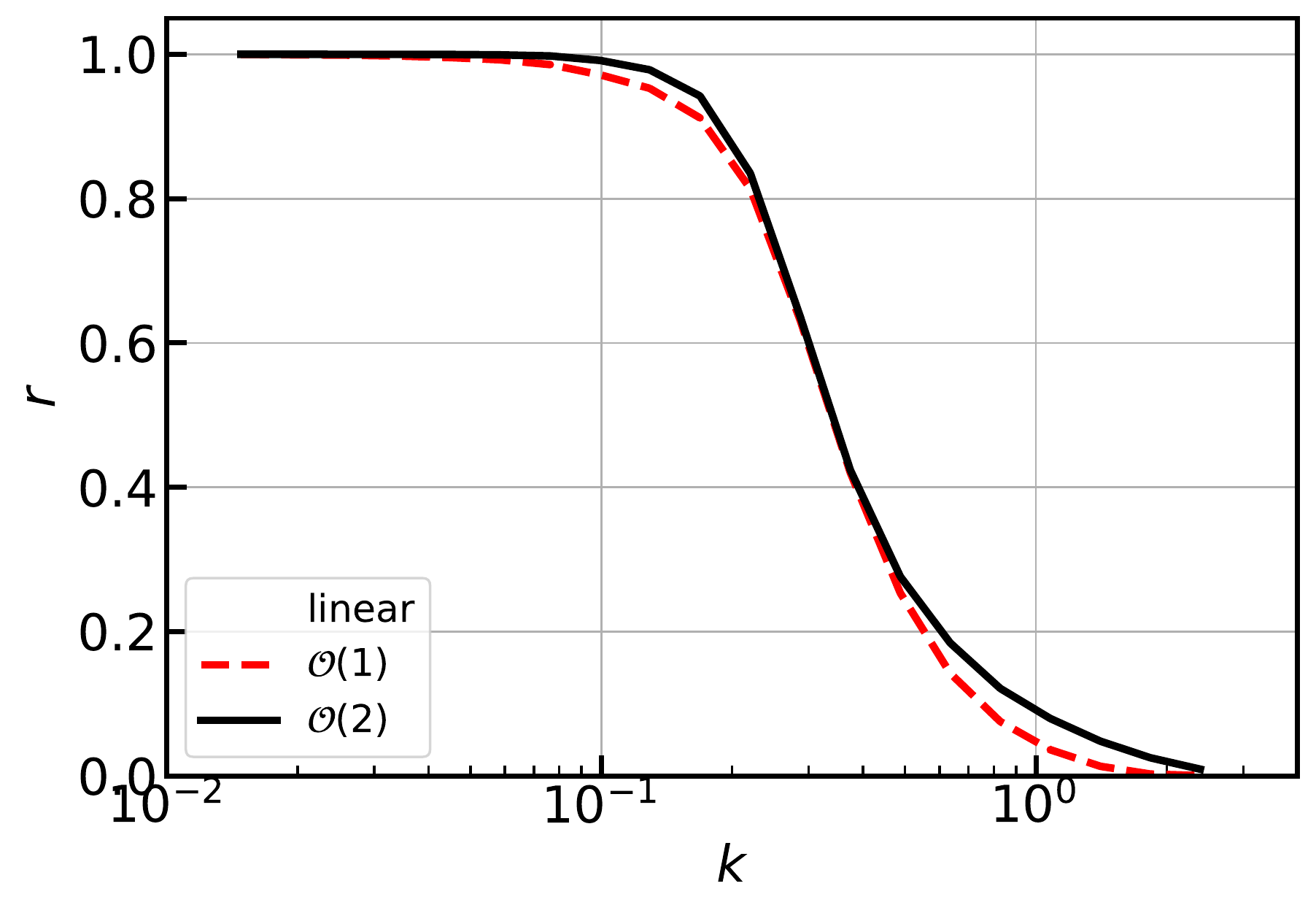}\\
\plotone{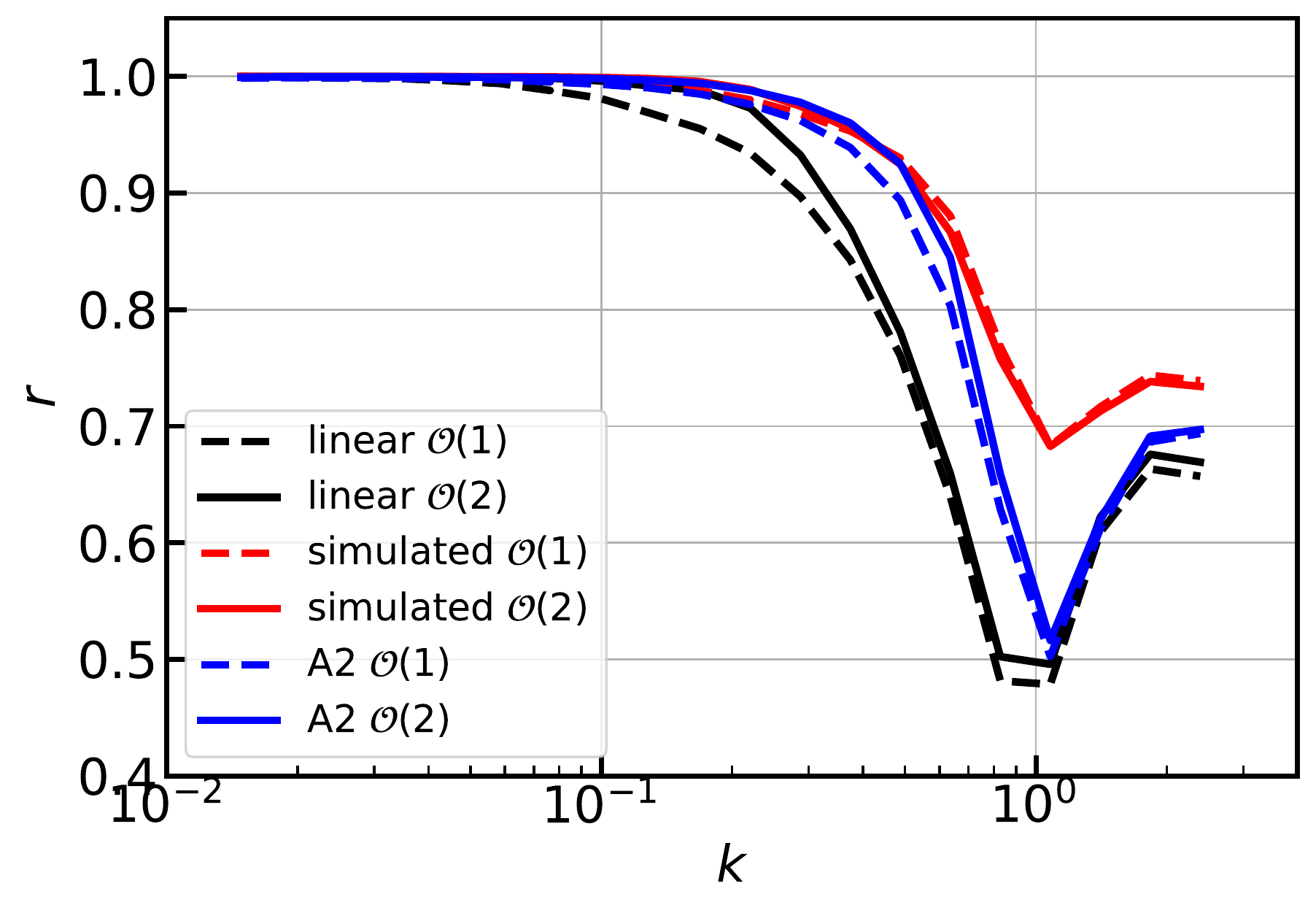}
\caption{The $\mathcal{O}(1)$ and $\mathcal{O}(2)$ reconstruction using the linear density.
The top panel is the cross-correlation in Lagrangian space.
The bottom panel is the result in Eulerian space.
Also plotted in the bottom panel is the cross-correlation result using the simulated and the reconstructed displacement for comparison.
\label{fig:linearrecon}}
\end{figure}

For the purpose of mock construction, it is straightforward to start with a linear density field.
Here we test the performance of the velocity reconstruction using the linear density field.
Combined with the fast density map generation methods such as 1LPT, 2LPT, or other techniques, we can obtain mocks with both good density and velocity field.
These synthetic mocks are of great importance for future surveys.

The process is roughly the same as the reconstruction with the simulated displacement.
We just use the linear density field to replace the simulated displacement divergence in the reconstruction algorithm.
Since the linear density is Gaussian, no orthogonization is needed.

The transfer function in this case is measured by 
\begin{eqnarray}
\begin{aligned}
&T_1(k)=\frac{\langle\delta_L\theta_t\rangle}{\langle\delta_L\delta_L\rangle}\ ,\\
&\theta_r^{(1)}(\boldsymbol{k})=T_1(k)\delta_L(\boldsymbol{k})\ .
\end{aligned}
\end{eqnarray}
We could also perform $\mathcal{O}(2)$ reconstruction by further measuring the second transfer function from the second-order LPT density field $\delta^{(2)}$ and the residual velocity field $\theta_m=\theta_t-\theta_r^{(1)}$:
\begin{eqnarray}
\begin{aligned}
&T_2(k)=\frac{\langle\delta^{(2)}\theta_m\rangle}{\langle\delta^{(2)}\delta^{(2)}\rangle}\ ,\\
&\theta_r^{(2)}(\boldsymbol{k})=T_2(k)\delta^{(2)}(\boldsymbol{k})\ .
\end{aligned}
\end{eqnarray}
In this case, the $\mathcal{O}(2)$ reconstruction (i.e., 2LPT) captures more nonlinear velocity information than the $\mathcal{O}(1)$ reconstruction (i.e., 1LPT). 
This is observed in the top panel of Fig. \ref{fig:linearrecon}.
$\mathcal{O}(2)$ reconstruction increases the cross-correlation coefficient at $k\sim 0.1\ \hmpc$ and $k\sim 1\ \hmpc$ in Lagrangian space.

In the bottom panel, we compare the velocity cross-correlation coefficient in Eulerian space with the reconstruction using the linear displacement (black lines), using the simulated nonlinear displacement (red lines), and using the reconstructed displacement (blue lines).
The $\mathcal{O}(1)$ reconstruction is presented as a dashed line and the $\mathcal{O}(2)$ reconstruction is a solid line.
From this plot we clearly see that the case using the nonlinear or reconstructed displacement produces a better performance than the case using linear displacement.
We also note that the velocity reconstruction performance using the reconstructed displacement catches the upper limit down to scale $k\sim 0.7\ \hmpc$.

\subsection{Improvement by real space transfer functions}

\begin{figure}[ht!]
\plotone{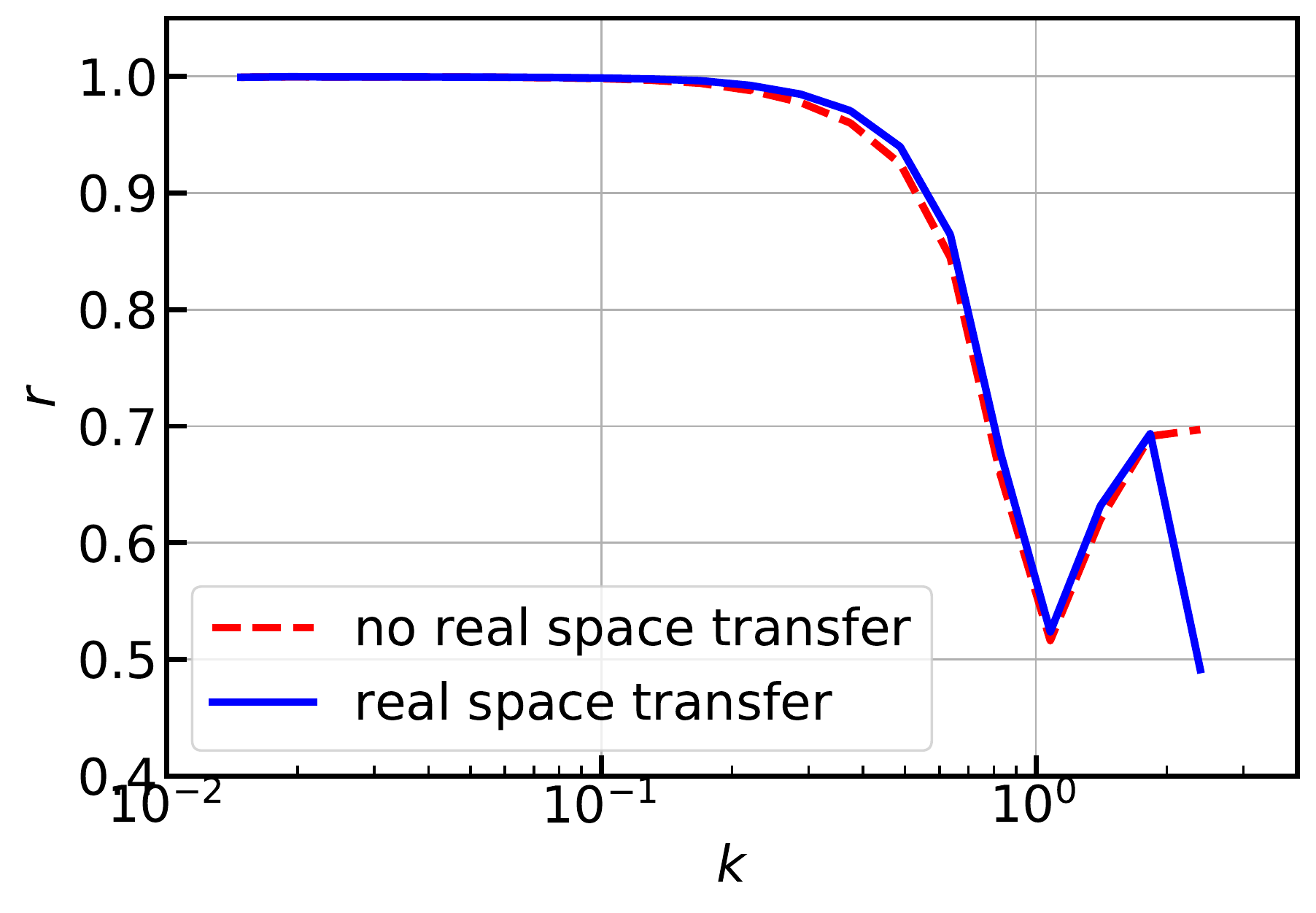}
\caption{The red dashed line presents the cross-correlation coefficient between the reconstructed velocity and the true velocity field.
The blue solid line is the result of two real space transfer functions being applied to make an optimal combination of the E-mode and B-mode components.
A very mildly improvement is observed.
\label{fig:EB-transfer}}
\end{figure}

According to the result in Section \ref{sec:bmode-in}, we should only use the E-mode displacement in the reconstruction, and the reconstructed velocity field is irrotational in Lagrangian space by design.
However, after mapping to Eulerian space, a part of the E-mode is converted into B-mode.
To obtain a better velocity reconstruction in real space, we could measure two more transfer functions in Eulerian space to adjust the reconstructed E- and B-mode component:
\begin{eqnarray}
\begin{aligned}
&T^E(k)=\frac{\langle{v}^E_r{v}^E_t\rangle}
{\langle{v}^E_r{v}^E_r\rangle}\ ,\\
&T^B(k)=\frac{\langle{v}^B_r{v}^B_t\rangle}
{\langle{v}^B_r{v}^B_r\rangle}\ ,
\end{aligned}
\end{eqnarray}
and $\boldsymbol{v}_r(\boldsymbol{k})=T^E(k)\boldsymbol{v}^E_r(\boldsymbol{k})+T^B(k)\boldsymbol{v}^B_r(\boldsymbol{k})$.
The result is presented in Fig. \ref{fig:EB-transfer}.
A very mild improvement is observed in the cross-correlation coefficient at $k\sim 0.5\ \hmpc$.
Considering that the improvement is negligible and this process may induce noise instead of improving the performance if the transfer function is not sufficiently accurate, we do not propose to applying this final step in the reconstruction.

\section{Conclusion and Discussion}
\label{sec:conclusion}

\begin{table*}[ht!]
\centering
\caption{A summary of the performance from different reconstruction methods, including the standard reconstruction, the $\mathcal{O}(1)$ and $\mathcal{O}(2)$ nonlinear reconstructions by three recently developed algorithms, the reconstruction using the simulated displacement, and with the linear density field.} \label{tab:summary}
\begin{tabular}{c|ccccccc}
\tablewidth{0pt}
\hline
\hline
  &
$\mu$ &
$\Delta v$ & $\Delta v$ &
$v_t-v_p$ & $v_t-v_p$ &
$\Delta v_z$ & $\Delta v_z$ \\
 & & Mean & rms  & Mean & rms  & Mean & rms \\
\hline
Standard & 0.958 & 22.3 & 71.5 & 33.6 & 75.8 & 0.02 & 72.5  \\
\hline
{A1} $\mathcal{O}(1)$ & 0.965 & -6.34 & 71.8 & 3.25 & 77.9 & -17.6 & 62.3  \\
{A1} $\mathcal{O}(2)$ & 0.968 & 10.2 & 64.2 & 18.7 & 69.6 & -16.1 & 57.8 \\
{A2} $\mathcal{O}(1)$ & 0.971 & 5.10 & 67.3 & 13.4 & 72.9 & 0.08 & 61.4 \\
{A2} $\mathcal{O}(2)$ & 0.977 & 23.5 & 57.5 & 29.9 & 62.7 & 0.07 & 55.4 \\
{A3} $\mathcal{O}(1)$ & 0.970 & 5.71 & 70.7 & 14.2 & 78.2 & 9.19 & 63.7 \\
{A3} $\mathcal{O}(2)$ & 0.972 & 19.9 & 63.9 & 27.7 & 70.5 & 9.75 & 60.7 \\
\hline
Simulated $\mathcal{O}(1)$ & 0.978 & 10.9 & 62.1 & 17.4 & 69.3 & 0.04 & 56.5 \\
Simulated $\mathcal{O}(2)$ & 0.979 & 17.2 & 56.0 & 23.3 & 63.3 & 0.05 & 54.0 \\
Linear $\mathcal{O}(1)$ & 0.955 & 6.25 & 88.8 & 18.9 & 95.9 & 0.02 & 77.9 \\
Linear $\mathcal{O}(2)$ & 0.963 & 20.2 & 70.9 & 30.2 & 78.1 & 0.10 & 67.4 \\
\hline
\end{tabular}
\end{table*}

We propose a new velocity reconstruction method based on the estimated displacement field from the nonlinear density maps by recently developed algorithms.
The reconstruction is first performed in Lagrangian space by the calibrated transfer functions, and then the Eulerian velocity is obtained by the mapping.
We found that this new velocity reconstruction has a better performance than the standard reconstruction method based on the linearized continuity equation.  
It produces a velocity field with a better cross-correlation coefficient, less velocity misalignment, and smaller amplitude difference with the true one.
Generally, $\mathcal{O}(2)$ reconstruction outperforms $\mathcal{O}(1)$ reconstruction by taking use of the velocity information residing in the high-order terms.
A summary of the statistics we investigated is presented in Table \ref{tab:summary}.

We explored several extensions.
One extension is to consider the correlation between the B-mode component of the velocity and displacement.
The other one adopts two more transfer functions in real space to adjust the reconstructed E-mode to the B-mode components in Eulerian space.
However, the performance is not improved or very mild.
Thus, it complicates the process and is not paid off. 

We also explored the upper limit of this new reconstruction method by assuming the displacement is perfectly reconstructed.
We found that in this case the difference between $\mathcal{O}(1)$ and $\mathcal{O}(2)$ is very small.
One surprising point is that the $\mathcal{O}(2)$ reconstruction performance from A2 is very close to this upper limit in the sense of the cross-correlation coefficient and the misalignment angle.
We also attempted to obtain the velocity field from the linear displacement with the same approach.
This demonstrates the limit of only using the 1LPT and 2LPT displacements and the transfer functions with calibration.

The reconstruction performance from a biased tracer depends on the understanding of the bias.
\cite{Wang19} found that the acoustic peaks are recovered best when the linear bias is correctly removed, and thus it is possible to obtain an estimation of the bias in the process of the reconstruction. 
For the low-density sample with only massive halos, correction of the bias is important.
Otherwise the overestimation of the displacement amplitude significantly degrades the linear density reconstruction.
For high-density sample with bias less than unity, the bias does not influence the results much  (e.g. \cite{Birkin19}).
We leave the quantification of the velocity reconstruction performance from the biased tracer to future investigations.

With the reconstructed initial condition of some volume of the universe, one could perform the simulation to obtain the velocity field (e.g., \cite{Lavaux08a}).
However, the reconstruction induces noise and non-Gaussianity in the linear density field.
A comparison between the constrained simulation and the original one has not been performed for the reconstruction algorithms investigated in this work.
We also leave the comparison of both the density and the velocity between the constraint simulation and the true one for the future.



\section*{Acknowledgement}
We thank Baojiu Li for providing the reconstruction codes and Pengjie Zhang for useful discussions.
This work was supported by the National Key Basic Research and Development Program of China (No. 2018YFA0404504), and
the National Science Foundation of China (grants No. 11773048, 11621303, 11403071).

\appendix

\section{Comparison of A1, A2, and A3}
\label{sec:appendix}

\begin{figure*}[ht!]
\plottwo{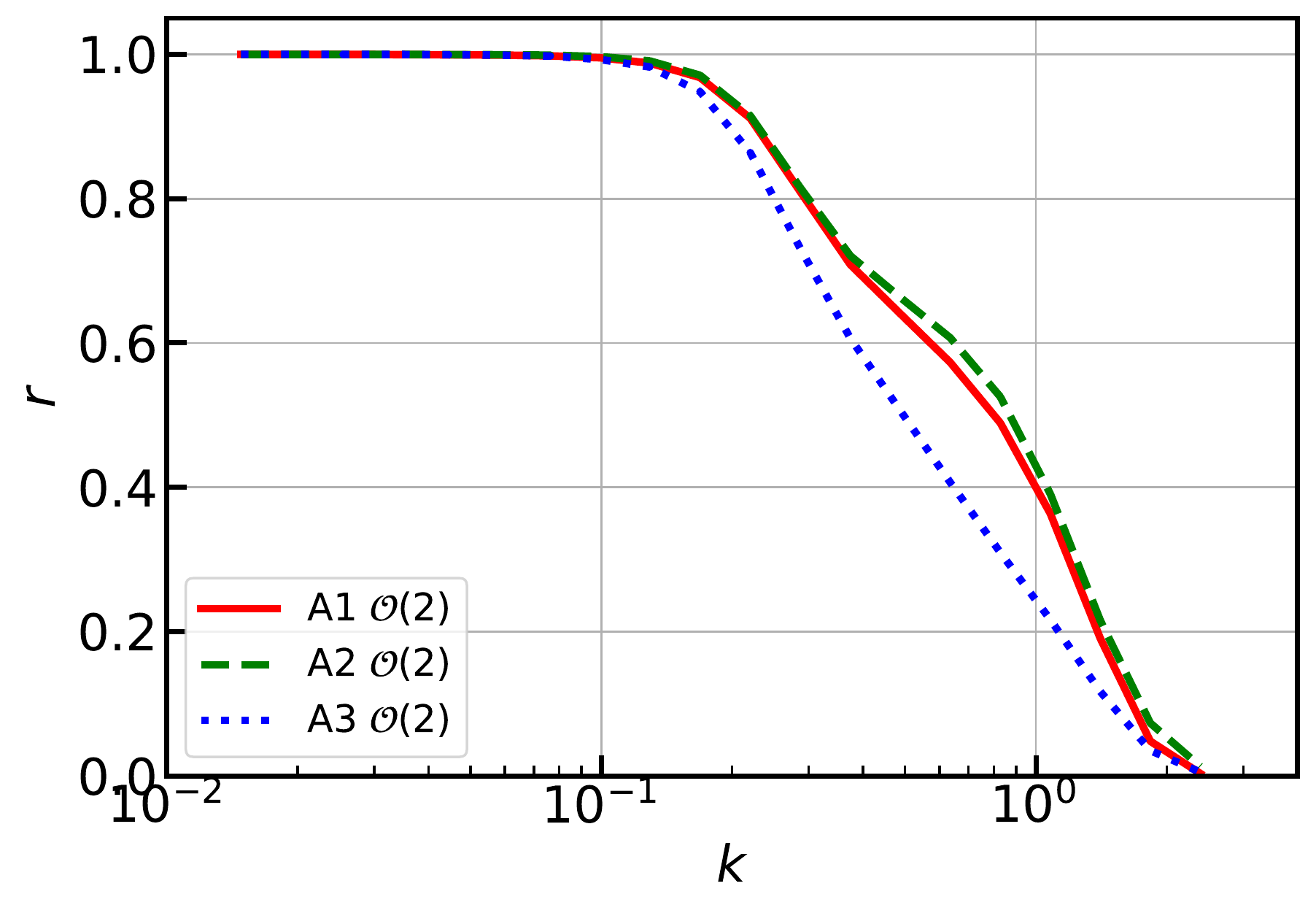}{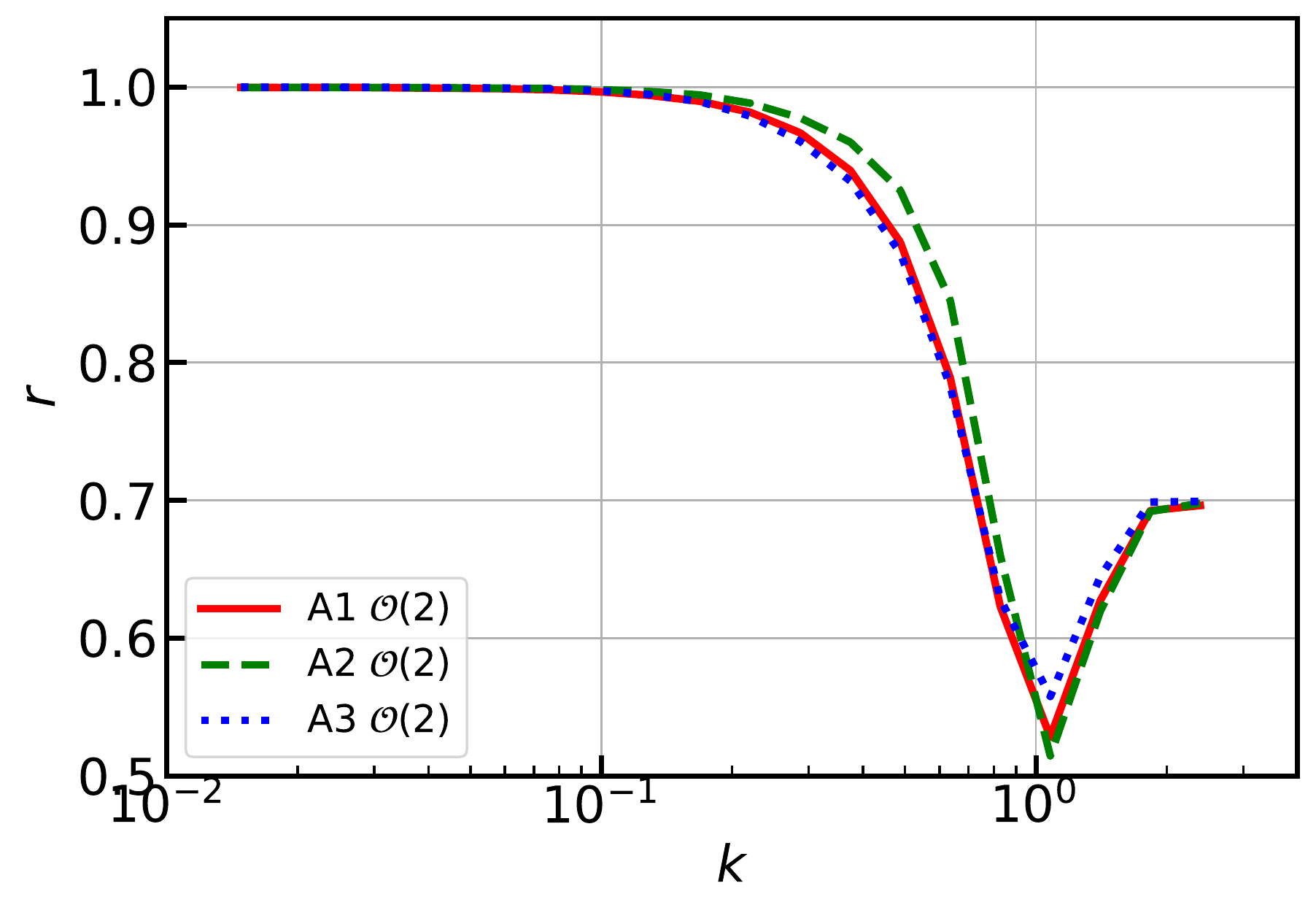}
\caption{Comparison of the three recently developed reconstruction algorithms.
The left panel shows the cross-correlation coefficients between the true velocity and the $\mathcal{O}(2)$ reconstructions in Lagrangian space and the right panel shows the coefficients in Eulerian space.
\label{fig:recon3}}
\end{figure*}

In this work we adopted the three algorithms A1, A2, and A3.  They are described in \cite{Zhu17}, \cite{Shi18}, \cite{Schmittfull17}, respectively.
These tests have not been performed in a systematical way before and a brief comparison is presented here.
We use the default parameters proposed in the literature since the simulation and analysis configuration is similar.
The result of the reconstructed Lagrangian velocity is presented in Fig. \ref{fig:recon3}.
The solid red, dashed green, and dotted blue lines are for A1, A2, and A3, respectively.
The left panel is the cross-correlation coefficient of the reconstructed and true velocities in Lagrangian space, while the right panel is for the Eulerian space.
We find the A2 has the best cross-correlation coefficient in Lagrangian space.
The good performance is also mapped into Eulerian space, leading to the best 
cross-correlation coefficient for Eulerian velocity.
However, A1 has a similar but slightly worse performance to A2 in Lagrangian space, and has a similar performance with A3 in Eulerian space.
A summary is presented in Table \ref{tab:summary}.


\bibliographystyle{aasjournal}
\bibliography{main}

\clearpage

\end{document}